\documentclass[aps,prb,twocolumn,showpacs,amsmath,amssymb]{revtex4-1}
\usepackage{graphicx}
\usepackage{color}
\usepackage[colorlinks,plainpages=false,linkcolor=black,urlcolor=blue,citecolor=blue,pdfpagemode=UseNone,pdfstartview=FitH]{hyperref}
\usepackage[T1]{fontenc}
\usepackage[ansinew]{inputenc}
\usepackage[english]{babel}
\usepackage[left=2.5cm,right=2.5cm,top=2cm,bottom=3.0cm,a4paper]{geometry}
\usepackage{tabularx}
\usepackage{color}
\usepackage{graphicx}
\usepackage{dcolumn}
\usepackage{bm}
\usepackage{bbold}
\usepackage{mathtools}

\newcommand{\bea}{\begin{eqnarray}}
\newcommand{\eea}{\end{eqnarray}}
\newcommand{\beq}{\begin{equation}}
\newcommand{\eeq}{\end{equation}}
\newcommand{\bd}{\begin{displaymath}}
\newcommand{\ed}{\end{displaymath}}
\newcommand{\ba}{\begin{array}}
\newcommand{\ea}{\end{array}}
\newcommand{\bi}{\begin{itemize}}
\newcommand{\ei}{\end{itemize}}
\newcommand{\bc}{\begin{center}}
\newcommand{\ec}{\end{center}}
\newcommand{\bfl}{\begin{flushleft}}
\newcommand{\efl}{\end{flushleft}}
\newcommand{\bfr}{\begin{flushright}}
\newcommand{\efr}{\end{flushright}}

\def\6{\partial}

\def\={\!\!\!&=&\!\!\!}
\def\+{\!\!\!&&\!\!\!+~}
\def\-{\!\!\!&&\!\!\!-~}

\newcolumntype{C}{@{}>{\hspace*{5mm}}c}

\newcommand{\RN}[1]{\text{\uppercase\expandafter{\romannumeral #1\relax}}}

\newcommand{\be}{\begin{equation}}
\newcommand{\ee}{\end{equation}}

\def\6{\partial}

\def\={\!\!\!&=&\!\!\!}
\def\+{\!\!\!&&\!\!\!+~}
\def\-{\!\!\!&&\!\!\!-~}

\begin{document}

\title[]{Quasiparticle interference from different impurities on the surface of  pyrochlore iridates: signatures of the Weyl phase}
\author {F. Lambert$^{1}$, A. P. Schnyder$^2$, R. Moessner$^3$, and I. Eremin$^1$}
\affiliation{
$^{1}$Institut f\"ur Theoretische Physik III, Ruhr-Universit\"at Bochum, D-44801 Bochum, Germany\\
$^{2}$Max-Planck-Institut f\"ur Festk\"orperforschung, Heisenbergstrasse 1, D-70569 Stuttgart, Germany \\
$^{3}$Max Planck Institute for the Physics of Complex Systems, D-01187 Dresden, Germany
}

\begin{abstract}
Weyl semimetals are gapless three-dimensional topological materials where two bands touch at
an even number of points in the bulk Brillouin zone. These semimetals exhibit topologically protected surface
Fermi arcs, which pairwise connect the projected bulk band touchings in the surface Brillouin zone. Here, we analyze the quasiparticle interference patterns of the Weyl phase when time-reversal symmetry is explicitly broken. We use a multi-band $d$-electron Hubbard Hamiltonian on a pyrochlore
lattice, relevant for the pyrochlore iridate R$_2$Ir$_2$O$_7$ (where R is a rare earth). Using exact diagonalization, we compute the surface spectrum and quasiparticle interference (QPI) patterns for various surface terminations and impurities. We show that the spin and orbital texture of the surface states can be inferred from the absence of
certain backscattering processes and from the symmetries of the QPI features for  non-magnetic and magnetic impurities. Furthermore, we show that the QPI patterns of the Weyl phase in pyrochlore iridates may exhibit additional interesting features that go beyond those found previously in TaAs.
\end{abstract}
\date{\today}

\maketitle

\section{Introduction}\label{sec:1_introduction}

Topological invariants in condensed-matter systems  are defined on closed manifolds in
momentum space. For three-dimensional systems,
an important closed manifold in momentum space is a two-dimensional Fermi surface. In this regard one can define the so-called topological
metal by the Chern numbers of the single particle
wave functions at the Fermi surface energies. These nonzero Chern numbers arise when the Fermi surface
encloses a band-crossing point, referred as the Weyl point, which is a singular point of the Berry curvature in momentum space\cite{Fang2003,Wan2011,Balents2011,Xu2011}. Materials
with such Weyl points near the Fermi level are called Weyl
semimetals \cite{Nielsen1983}.

The Weyl semimetal state was proposed to be realized in Rn$_2$Ir$_2$O$_7$ pyrochlore systems 
with an all-in/all-out magnetic structure\cite{Wan2011} and in the ferromagnetic phase of
HgCr$_2$Se$_4$\cite{Xu2011}. Another proposal involved a fine-tuned
multilayer structure of normal insulators and magnetically doped topological insulators\cite{Burkov2011}.  Very recently, the Weyl semimetal phase was proposed in the Dirac semimetals Cd$_3$As$_2$\cite{Jeon2014} and Na$_3$Bi\cite{xiong_ong_Na3Bi_chiral_anomaly_science_2015} with an external magnetic field
applied along the axis of the Dirac points, 
and in the inversion-symmetry breaking systems TaAs\cite{huang_hasan_TaAs_nat_commun_15}, NbAs\cite{xu_hasan_NbAs_NatPhys_15}, TaP, and NbP\cite{Weng2015}.
For any lattice model, the Weyl points appear in pairs of opposite
chirality or monopole charge.  The only way to annihilate a
pair of Weyl points with opposite chirality is to move them
to the same point in the BZ. In this sense semimetals are topologically stable.
Important for surface sensitive experiments is that the existence of Weyl points near the Fermi level leads
to several unique physical properties, one of them is the appearance
of discontinuous Fermi surfaces (Fermi arcs) on the
surface\cite{Balents2011,Burkov2011,Weng2015}, which were confirmed experimentally in TaAs by ARPES\cite{Xu2015,Lv2015}. Photoemission measurements have observed conical dispersions away from certain points
in the Brillouin zone of these materials.

In order to
analyze the physics near the Weyl points and to clarify the effects of
material inhomogeneities on the low-energy behaviour,  high energy-resolution,
atomically resolved spectroscopic measurements are important. In this regard the use of low-temperature
Fourier transformed scanning tunnelling microscopy (FT-STM) is ideally suited to address these crucial issues, as was recently shown for TaAs\cite{Inoue2016}.
FT-STM measures the wave-length of Friedel oscillations caused by disorder present
in a metallic system, which in turn contains information on the electronic structure of
the pure system.  The wavelengths of these Friedel oscillations appear in the Fourier transformed STM data
as peaks at particular wavevectors {\bf q}, which disperse
with STM bias. In general there exists no
exact theoretical description for the \emph{intensities} of the QPI patterns,
since theses depend on the form of the impurity potentials and
on the {\bf k}-dependence of the tunneling matrix elements, 
which are in most cases unknown.
However, the \emph{positions} of the peaks in the QPI patterns do not depend
on these effects and are determined only by the electronic
structure of the pure system.

Most recently, the QPI patterns on the surface of inversion-symmetry breaking Weyl semimetals were analyzed  theoretically\cite{Mitchell2016,Kourtis2016}. Here, we extend these studies to inversion symmetric, time-reversal breaking Weyl semimetals, 
by considering an interacting multi-band $d$-electron Hubbard Hamiltonian on the pyrochlore
lattice with the antiferromangetic spin configuration of R$_2$Ir$_2$O$_7$.  We compute the surface spectrum and QPI patterns for this Weyl phase for different types of surface impurities and surface terminations using the $T$-matrix approximation. We demonstrate that the QPI patterns for the Weyl phase without time-reversal symmetry show unique features and lack the so-called ``pinch-point" at \({\bf q} = 0\) that was argued to be characteristic for the QPI of Weyl semimetals without inversion symmetry \cite{Kourtis2016}. In the present case this structure is completely suppressed, due to the non-trivial spin polarization of the surface states, but is  still perfectly visible in the joint density of states. Instead, we find as a clear signature of the Fermi arcs,  disjoint cross correlation arcs in the outer regions of the QPI patterns. These results can be used to  uniquely identify Weyl phases in condensed matter systems with time-reversal symmetry breaking.

The outline of the remainder of the paper is as follows: in Section \ref{tb_ham} we discuss the microscopic Hamiltonian\cite{pyro_ham} to describe the Weyl phase. We determine the mean-field phase diagram for this model including the antiferromagnetic all-in/all-out configuration at zero temperature and derive in Sec. III the  surface states for two kinds of surface terminations (triangular and kagome). In Sec.~IV we introduce the general procedure to calculate  the  QPI patterns in a slab geometry and present the results of these calculations.
The conclusions are presented in Sec.~V.

\section{Model of the Weyl phase in pyrochlore iridates}
\label{tb_ham}

\begin{figure}[t]
\centering
\includegraphics[width=1.0\linewidth]{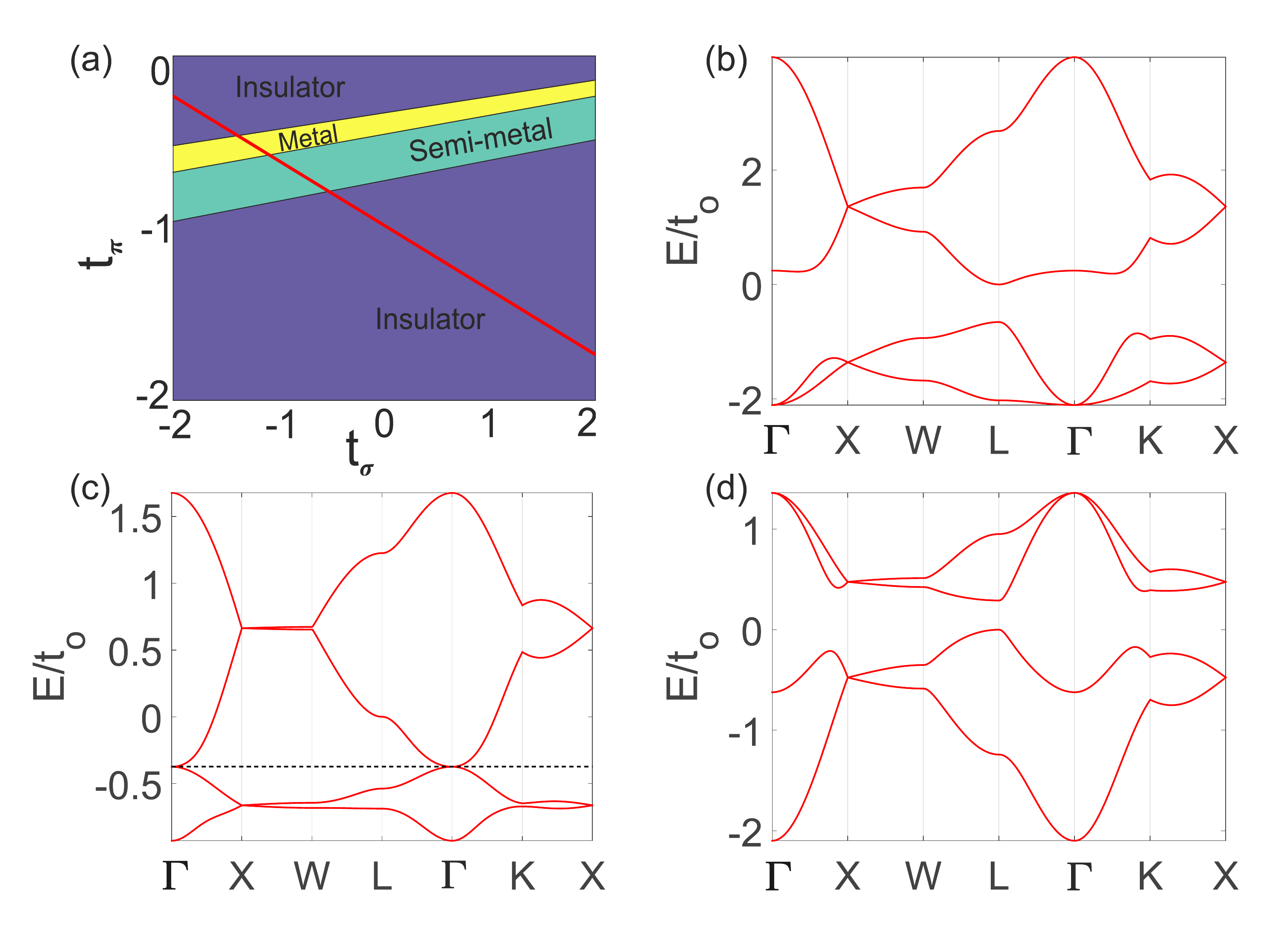}
\caption{ (a) Phase diagram of the non-interacting pyrochlore model\cite{tb_source,micro_pyro}, Eq.~(\ref{mom_ham}). The axes correspond to the values of the hopping parameters and the red line indicates the relation $t_{\pi} = - \frac{2}{3} t_{\sigma}$, used previously \cite{micro_pyro}. (b)-(d) Band structure for different values of $ t_{\sigma}=0.4t_o$ (b), $ t_{\sigma}=-0.8t_o$ (c), and $ t_{\sigma}=-2t_o$ (d). The dashed-dotted line in (c) refers to the position of the chemical potential for the quadratic bands touching.}
\label{fig7}
\end{figure}

Following the original proposal of the Weyl semimetal phase in pyrochlore iridates\cite{tb_source}, we study here an interacting tight-binding Hubbard Hamiltonian with hopping matrix elements along the \( \sigma \)- and \(\pi\)-bonds and along the oxygen-mediated bonds. The non-interacting part of the Hamiltonian including nearest- and next-nearest-neighbour hopping can be expressed in momentum space as:
 \begin{align} 
\label{mom_ham}
 & H^0_{\vec k} =  \sum_{a,b} \left[ H^{NN}_{ab} ({\vec k})+ H^{NNN}_{ab} ({\vec k}) \right] ,  \\
& H^{NN}_{ab} \left( \vec k \right) = 2 \left( t_1 +t_2 i \vec d_{ab} \cdot \vec \sigma \right) \cos \left( \vec k \cdot \vec b_{ab} \right), \notag\\
& H^{NNN}_{ab} \left( \vec k \right) = 2 \sum_{c \ne a,b} \{ t_1 \left( 1 - \delta_{ab} \right) +i \left[ t_2' \left( \vec b_{ac} \times \vec b_{cb} \right)   \right. \notag \\
& \quad + \left. t_3' \left( \vec d_{ac} \times \vec d_{cb} \right) \right] \cdot  \vec \sigma \} \cos \left( \vec k \cdot \left( - \vec b_{ac} + \vec b_{cb} \right) \right), \notag
\end{align}
where \( \vec b_{ij}\) is the vector connecting two corners of the tetrahedron and \( \vec d_{ij}= 2 \vec a_{ij} \times \vec b_{ij}\) is orthogonal to \( \vec b_{ij}\) and \(\vec a_{ij}\) pointing from the center of the tetrahedron to the connection of the corners \( i \) and \(j\). The positions of the Iridium atoms are \( b_1 = \left( 0,0,0 \right), b_2 = \left( 0,1,1 \right), b_3 = \left( 1,0,1 \right)$, and $b_4 = \left( 1,1,0 \right)\) in this coordinate frame.
By comparing the model in a local description \cite{micro_pyro} with Eq.~(\ref{mom_ham}) one finds  expressions for \( \left( t_1 , t_2 , t_1' , t_2' , t_3' \right) \)  in terms of the hopping parameters (\( t_{\pi},t_{\pi}',t_{\sigma},t_{\sigma}',t_{o}\)) resulting from the \( \sigma \)- ,\(\pi\)- and an oxygen-mediated bonds. The values are given by
\begin{equation}
\begin{split}
t_1&= \frac{130}{243} t_o +\frac{17}{243} t_{\sigma} - \frac{79}{243} t_{\pi},\\
t_2&= \frac{28}{243} t_o +\frac{15}{243} t_{\sigma} - \frac{40}{243} t_{\pi},\\
t_1'&= \frac{233}{2916} t_{\sigma}' -\frac{407}{2187} t_{\pi}',\\
t_2'&= \frac{2}{2916} t_{\sigma}' -\frac{220}{2187} t_{\pi}' , \\
t_3'&= \frac{50}{2916} t_{\sigma}' -\frac{460}{2187} t_{\pi}' ,
\end{split}
\end{equation}
where \( t_{\sigma,\pi}'\) are the next-nearest-neighbour hopping parameters. This model of non-interacting fermions  exhibits three different phases\cite{tb_source,micro_pyro} as shown Fig.~\ref{fig7}.

\begin{figure}[]
 \includegraphics[width=\linewidth]{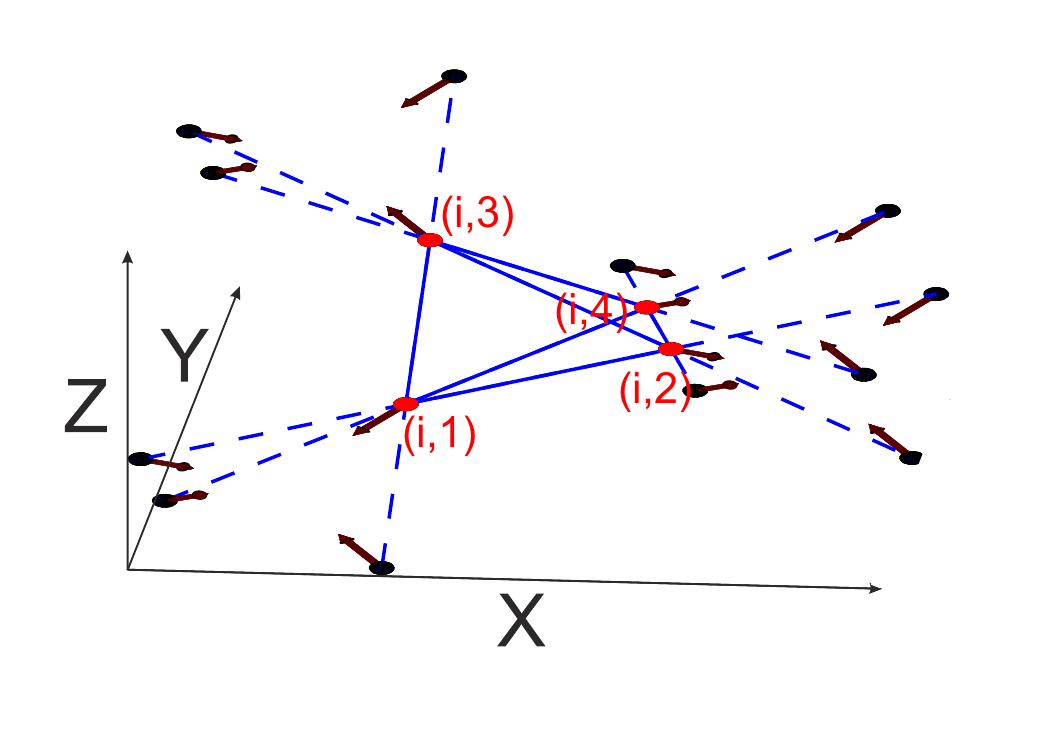}
     \caption{The unit cell of the pyrochlore lattice showing the magnetic all-in or all-out configuration of the magnetic moments.}
\label{fig8}
\end{figure}

\begin{figure}[b!]
    \includegraphics[width=1.0\linewidth]{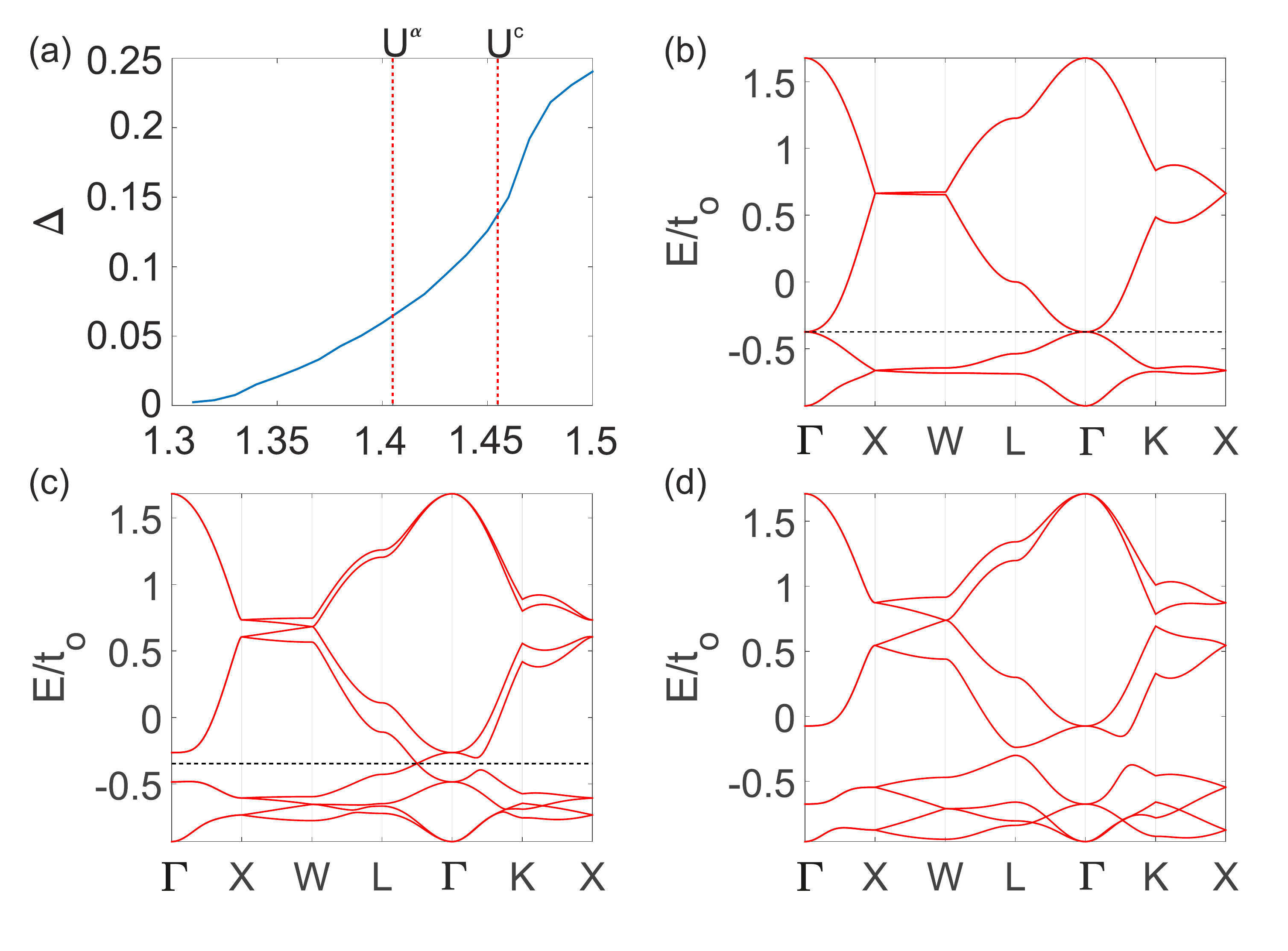}
  \caption{(a) Evolution of the magnetic moment with the strength of the Hubbard interaction $U$, computed for  $t_{\sigma}=-0.8 t_o$.  (b), (c) Corresponding band structures for different values of $U=1.53 t_o$ (b) and $U=1.6 t_o$ (c). Note that in the case of (c), the band structure exhibits the linear band crossings with a Weyl point, shown by the dashed-dotted line.}
\label{fig9}
\end{figure}

\begin{figure}[]
 \includegraphics[width=\linewidth]{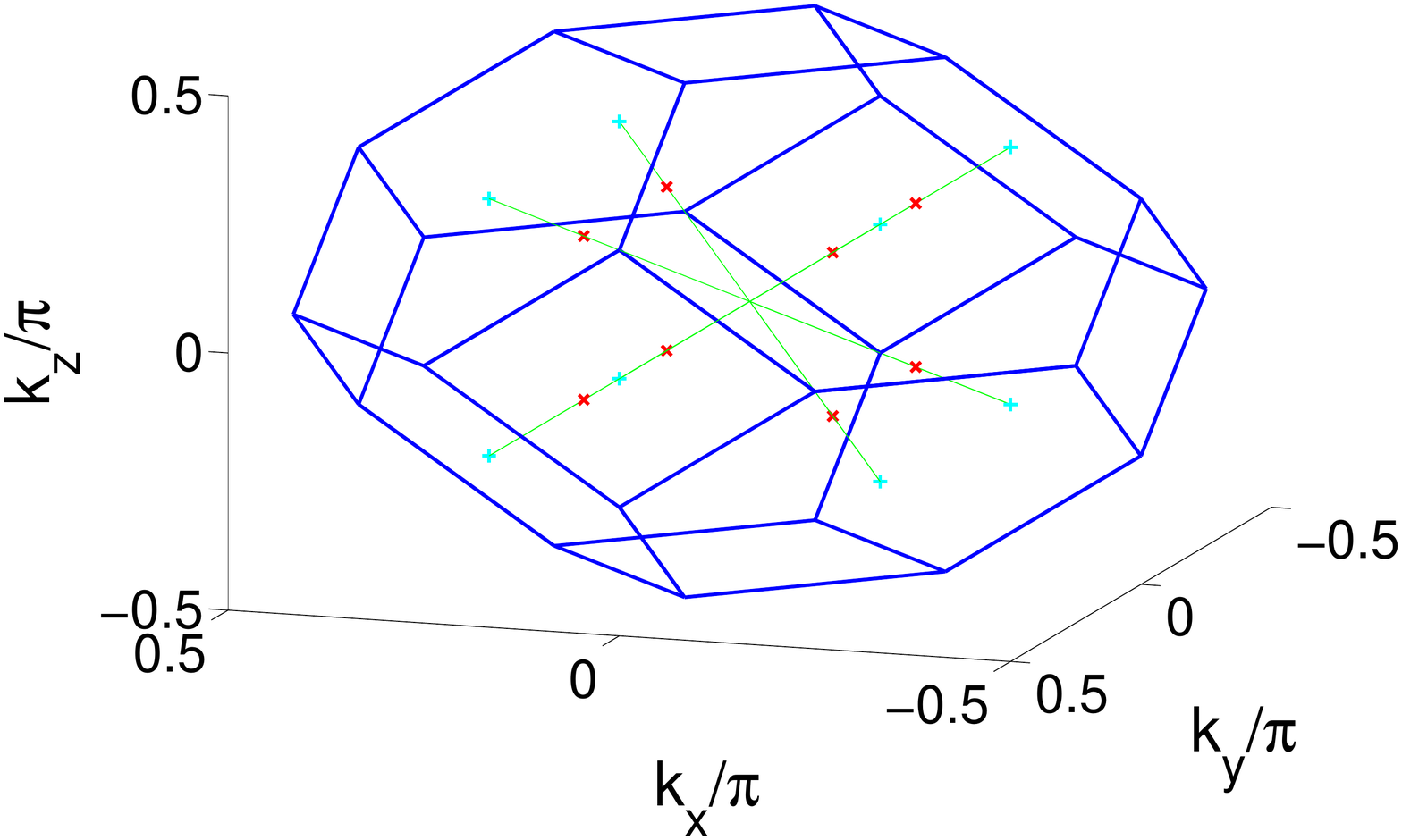}
     \caption{First Brillouin zone of the pyrochlore iridates with high symmetry ($\Gamma - L$) lines shown in green. Positions of the Weyl points and the $L$-points are shown by red and cyan crosses, respectively. Upon increasing the interaction, the Weyl points move along the symmetry lines and annihilate at the $L$-points .}
\label{fig10}
\end{figure}

At the next step we consider the effect of the on-site Hubbard like interaction for the metallic phase shown in Fig.~\ref{fig7}(c), which without magnetic order possesses time-reversal and inversion-symmetry. To account for the magnetic order consider the mean-field decoupling of the Hubbard interaction in the form:
\begin{align}
H^U&=U \sum_i n_{i \uparrow} n_{i \downarrow} \\
& \rightarrow -U \sum_i \sum_l \left( 2 \langle \vec j_{i l} \rangle \cdot \vec j_{i l} - \langle \vec j_{i l} \rangle^2 \right),
\label{mean_field_inter}
\end{align}
where \(i\) indicates the unit cell, while \(l=1\ldots4\) refers to the sites of the unit cell. Here, \(\vec j_{i l}=\sum_{\sigma \sigma'}\frac{ c_{l \sigma}^{\dagger} \vec \sigma_{\sigma \sigma'} c_{l \sigma'}}{2}\) are the local spin operators, whose expectation value are computed self-consistently  with respect to a specific magnetic configuration that preserves the unit cell size with the  antiferromagnetic phase, shown in Fig.~\ref{fig8}. Here, \( \langle \vec j_{i l} \rangle = \langle \vec j_{l} \rangle = \Delta_l = \Delta\) is the magnetic moment at each corner of the tetrahedron.
Fig.~\ref{fig9} shows the result of the numerical calculations with increasing strength of the Hubbard interaction, which agrees with the results of  Ref.~\onlinecite{tb_source}.
\begin{table}
\begin{tabular}[c]{|c|c|c|}\hline
  $\mathrm {Position} \left[ k_w \right]$ & $C_1$ \\ \hline
  $ \left( 1 , 1 , 1 \right)$ & 1 \\
  $\left( 1 , 1 , -1 \right)$ & -1 \\
  $\left( 1 , -1 , -1 \right)$ & 1 \\
  $\left( -1 , -1 , -1 \right)$ & -1 \\
  $\left( -1 , -1 , 1 \right)$ & 1 \\
  $\left( -1 , 1 , 1 \right)$ & -1 \\
  $\left( -1 , 1 , -1 \right)$ & -1 \\
  $\left( 1 , -1 , 1 \right)$ & 1 \\ \hline
\end{tabular}
\caption{Chern numbers of the Weyl points, shown in Fig.~\ref{fig10} by the red dots.}
\label{table1}
\end{table}
Most importantly for the intermediate value of $U$, as shown in Fig.~\ref{fig9}(c),  the band structure in the antiferromagnetic phase exhibits a linearly dispersing band touching along the \(L-\Gamma\)-line. The complete 3D Brillouin zone includes 8 such band touchings, as shown in Fig.~\ref{fig10}, each of them carrying a topological charge, as can be found by computing the Chern number from the Berry curvature, see Table I.

\section{Surface States}

\begin{figure}[t!]
 \includegraphics[width=\linewidth]{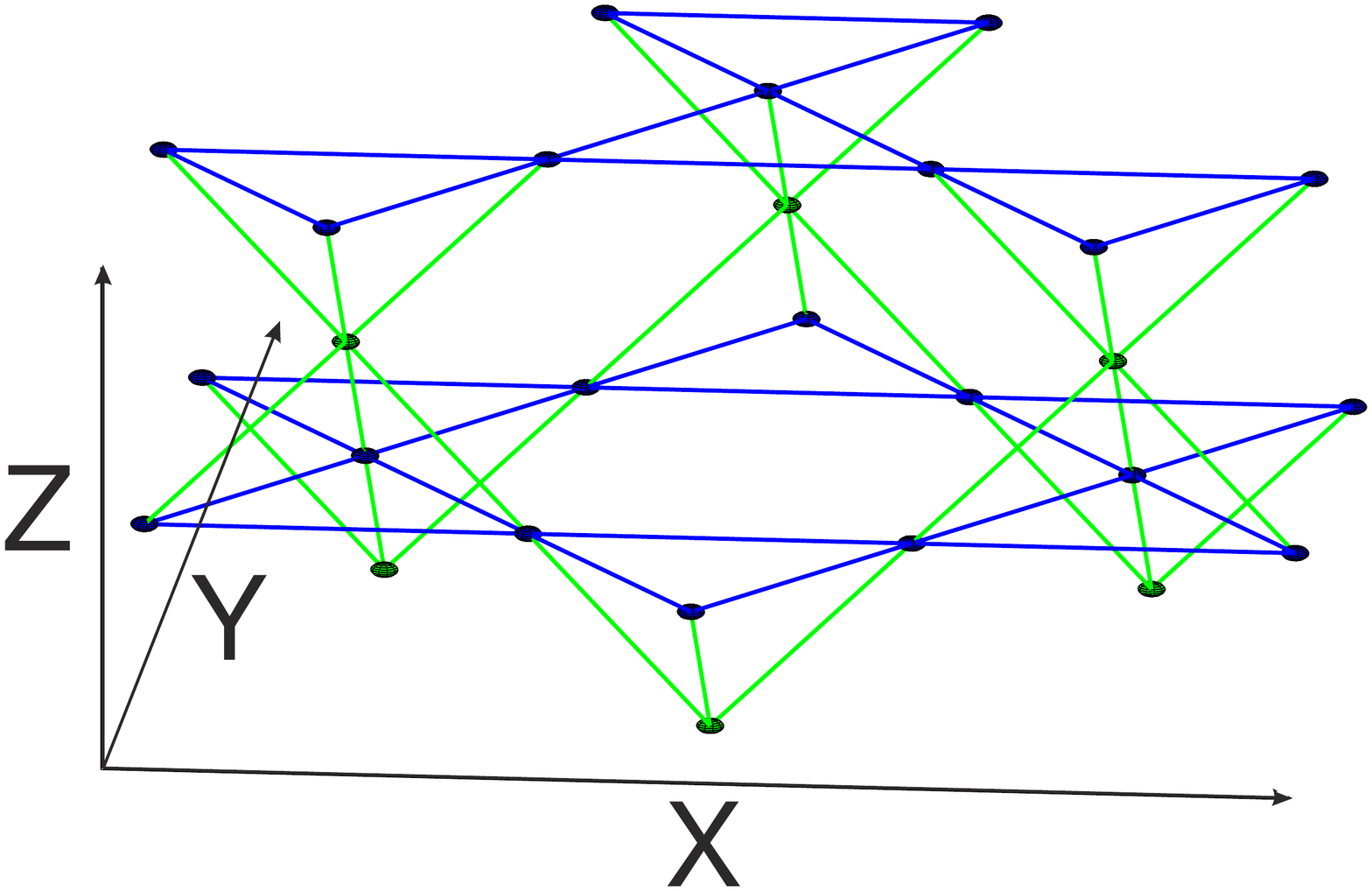}
     \caption{Lattice structure viewed in a new coordinate frame used for the various surface terminations.}
\label{fig11}
\end{figure}

\begin{figure}[b!]
\centering
   \includegraphics[width=0.9\linewidth]{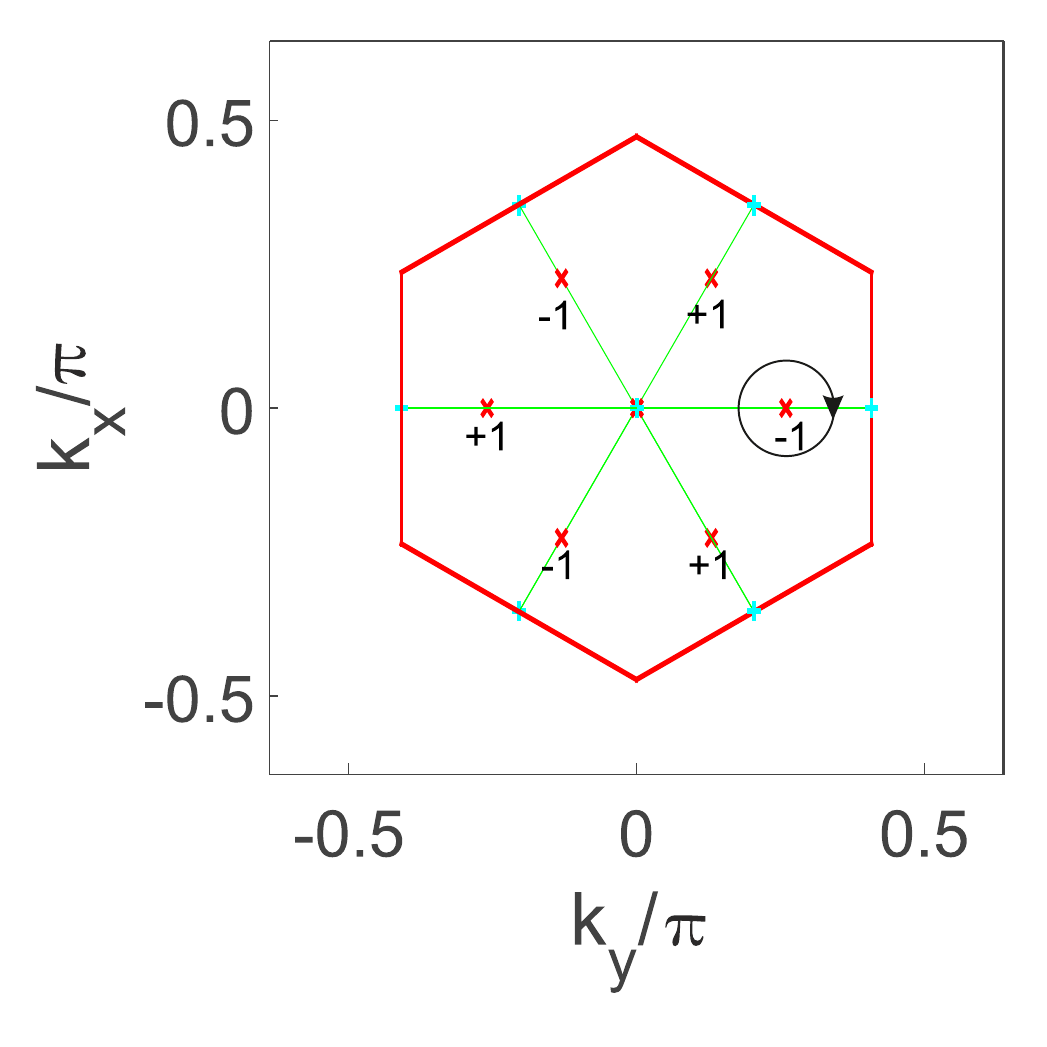}
    \caption{Projected Brillouin zone with the positions of the Weyl points (red crosses) for $U=1.41 t_o$ and their line of movement upon increasing the interaction strength (green lines). The lines terminate at the projected L-points (blue crosses), where the Weyl-Points meet up with those of the next Brillouin zone and annihilate. The black circle denotes the choice of the path (parametrized by the angle $\phi$) used in Fig.~\protect\ref{Afig12}.}
    \label{fig13a}
  \end{figure}

To perform the slab geometry calculations in the case of the pyrochlore lattice, it is important to find a suitable surface termination. Observe that the three-dimensional crystal lattice can be treated as a ``heterostructure'' of two two-dimensional sublattices, kagome and triangular ones being alternatingly stacked  as shown in Fig.~\ref{fig11}.  Re-parameterizing Eq.~(\ref{mom_ham}) with regard to the new coordinate frame we choose the $z$-direction as the stacking direction. The spin orientations are left in the old coordinate frame for simplicity, which is taken into account by suitably rotating the spin operator.
The 2D projected Brillouin zone of the surface is shown in Fig.~\ref{fig13a} together with the projected positions of the band touchings. For convenience the interaction strength is chosen in such a way that the Weyl points are located halfway between the \( \Gamma\)- and the L-points, which corresponds to \( U^{\alpha}\approx 1.41 t_o \). Note that since the Weyl points annihilate at  \( U^c\approx 1.46\), there exists a narrow region of stability of the Weyl phase.

\begin{figure}[t!]
\includegraphics[width=0.5\textwidth]{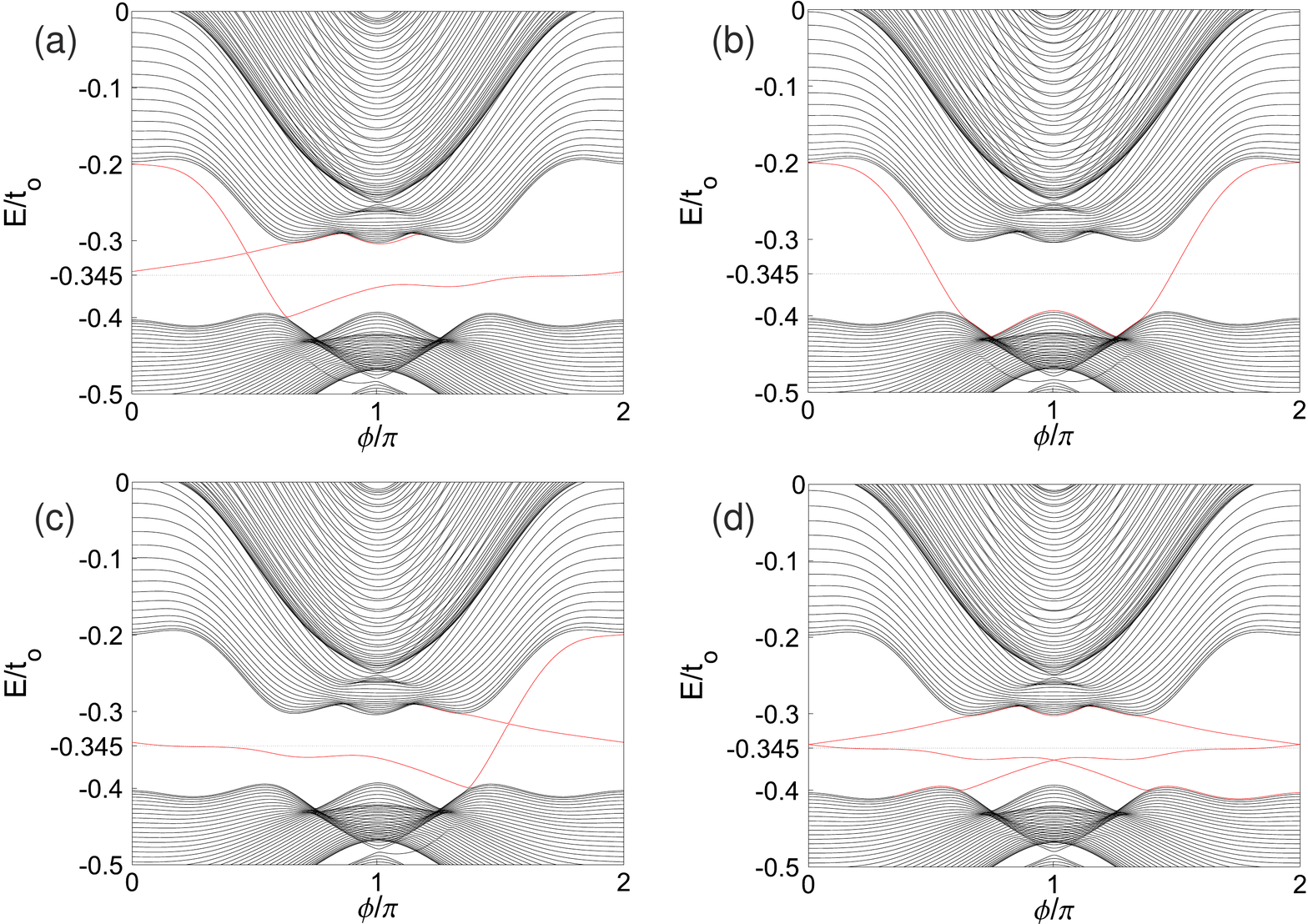}
\caption{Electronic dispersion in the Weyl phase of the pyrochlore iridates, calculated in the slab geometry around the Weyl point of $-1$ chirality with kagome-triangular (a), kagome-kagome (b), triangular-kagome (c), and triangular-triangular (d) surface terminations. The surface states are highlighted in red. Since the Fermi arcs always connect pairs of Weyl points, we plot the dispersion along a circle, shown in Fig.~\ref{fig13a}, enclosing one of the Weyl points.}
\label{Afig12}
\end{figure}

The submatrices corresponding to the hoppings within each layer will be indicated as \(H^0_{k}\) for the kagome lattice and \(H^0_{t}\) for the triangluar lattice. Inter-layer hopping matrices are indicated by \( H^{\pm i}_{ [t-k,k-t,k-k,t-t]}\) depending on which kinds of layers they connect and how many layers lie in between. From the Hamiltonian in Eq.~(\ref{mom_ham}) one finds that \(H^0_{k}\) is a \( 6 \times 6\) matrix, \(H^0_{t}\) a \( 2 \times 2\) matrix and the submatrices that contain the inter-layer hoppings are accordingly either $6 \times 2$, $2 \times 6$, or the same size as the onsite matrices. Starting with a kagome-lattice and finishing with a sparse triangular one, therefore only using full original unit cells, the overall Hamiltonian in the slab-geometry has the following form
\begin{eqnarray}
&& H_S \left( k_{||} \right) 
=
\\
&& \tiny{
\begin{pmatrix}
 H^0_k & H^{+1}_{k-t} & H^{+2}_{k-k} & H^{+3}_{k-t} & \ldots &  H^{+N-1}_{k-k} & H^{+N}_{k-t} \\
 H^{-1}_{t-k} & H^0_t & H^{+1}_{t-k} & H^{+2}_{t-t} & \ldots &  H^{+N-2}_{t-k} &  H^{+N-1}_{t-t} \\
 H^{-2}_{k-k} & H^{-1}_{k-t} & H^0_k & H^{+1}_{k-t} & \ldots  & H^{+N-3}_{k-k} &  H^{+N-2}_{k-t} \\
H^{-3}_{t-k} & H^{-2}_{t-t} & H^{-1}_{t-k} & H^0_t & \ldots  & H^{+N-4}_{t-k} &  H^{+N-3}_{t-t} \\
\vdots & \vdots & \vdots &  \vdots & \ddots &  & \vdots \\
 H^{-N+1}_{k-k} &  H^{-N+2}_{k-t} &  H^{-N+3}_{k-k} &  H^{-N+4}_{k-t} &   &  H^0_k  & H^{+1}_{k-t} \\
 H^{-N}_{t-k} &  H^{-N+1}_{t-t} &  H^{-N+2}_{t-k} &  H^{-N+3}_{t-t} & \ldots & H^{-1}_{t-k} & H^0_t
\end{pmatrix} }.
\nonumber
\label{slab}
\end{eqnarray}
%
To reveal the effects of different surface terminations on the band structure, the full dispersion including surface states is shown in Fig.~\ref{Afig12}  for  different combinations of surface terminations. Since the Fermi arcs always connect pairs of Weyl points, we plot the dispersion along a circle enclosing one of the Weyl points of \(-1\) chirality at some distance away from it, as shown by the black circle in Fig.~\ref{fig13a}. We find that there are two kinds of surface states corresponding to kagome  and triangular lattice terminations, respectively. The dispersions around the other Weyl point projections are either identical or related by a \(\phi \rightarrow - \phi\) transformation.

The Green's function in this slab representation is 
\begin{equation}
G \left( \vec k_{||} , i , j , \omega \right) = \sum_{n} \frac{ \psi_{\vec k_{||},i,n} \psi_{\vec k_{||},j,n}^{\dagger}}{\omega +i \eta - E_n \left(\vec k_{||} \right) },
\label{slab_greens_nd}
\end{equation}
where \(i,j\) are slab indices, \(n\) is the band index, and \(  \vec k_{||} \) is the in-plane momentum.  The slab Green's function can be numerically computed in an efficient way by matrix inversion of \(H_S\)
\begin{equation}
G_{i,j} \left( \vec k_{||} , \omega \right) = \left[ \frac{1}{\omega +i \eta - H_S \left(\vec k_{||} \right) } \right]_{i,j} .
\label{slab_greens_comp}
\end{equation}
\begin{figure*}[t!]
 \includegraphics[width=0.94\textwidth]{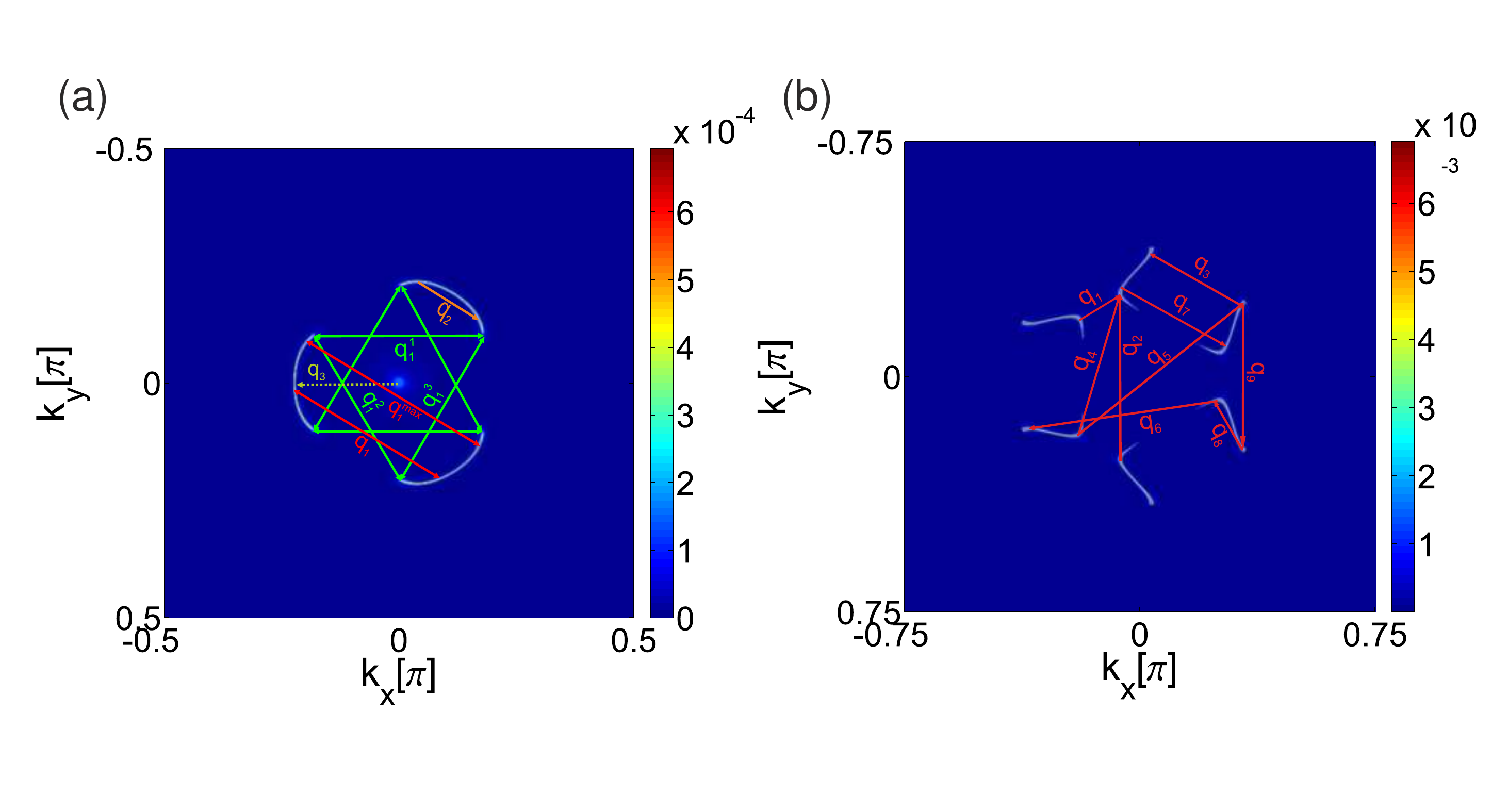}
     \caption{Spectral density of the kagome (a) and triangular (b) terminated surfaces for $\omega=-0345 t_o$ with possible scattering wave vectors introduced by impurities, as determined by the joint density of states, shown in Fig.~10.}
\label{fig14}
\end{figure*}
In Fig. \ref{fig14} we plot the resulting spectral densities for various terminated surfaces. As pointed out above, there are two possible regular surface terminations for the case of the iridates, having kagome and triangular lattice structures, respectively. Being the most natural choice, they also preserve the unit cells of the original model. For both terminations we find  Fermi arcs, connecting the two Weyl points of opposite chirality; see Fig. \ref{fig14}.
The choice of \( \omega \) in Fig. \ref{fig14} corresponds to the chemical potential at the surface. It is found to be slightly higher than that in the bulk by self-consistent evaluation in the slab-geometry basis. 

\begin{figure*}[t!]
    \includegraphics[width=0.94\linewidth]{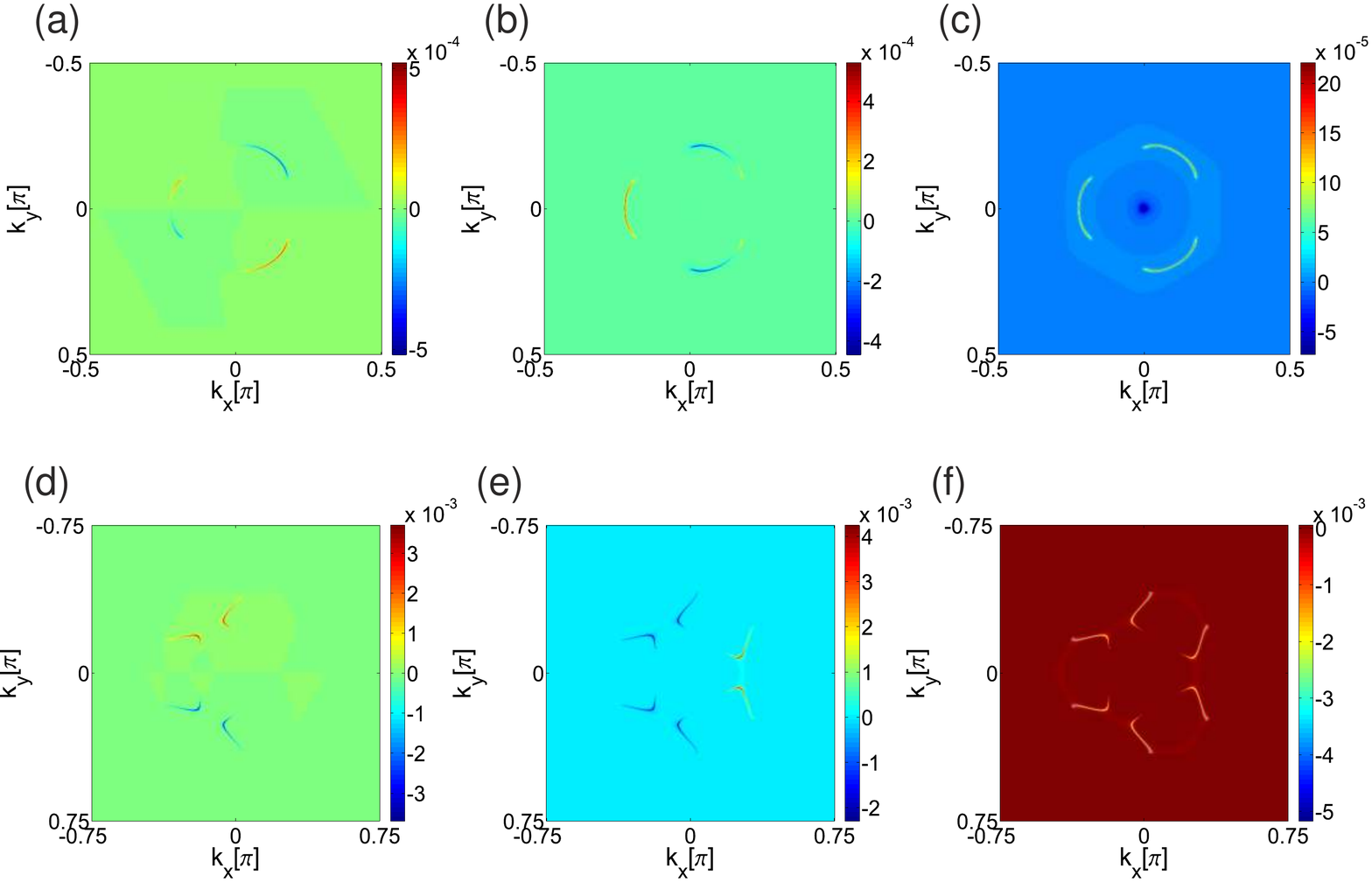}
\caption{ Calculated momentum map of the spin-resolved density of states for the surface states in the slab geometry for the kagome (upper panel)  and the triangular (lower panel) surface terminations for $\sigma_x$ [(a),(d)], $\sigma_y$ [(b),(e)], and $\sigma_z$ [(c),(f)] spin projections.}
\label{fig16}
\end{figure*}

To demonstrate the spin-momentum locking we plot in Fig.~\ref{fig16}  momentum maps of the spin-resolved density of states (SDOS) 
\begin{equation}  \label{DOS_spin-res}
\rho^{\alpha}_{\bf k}=-\frac{1}{\pi} \mathrm {Im}\left[ \mathrm {Tr} \left( \sigma^{\alpha} G_{11} (\bf k) \right) \right] .
\end{equation}
It is interesting to note that the spin-momentum locking of the surface states is non-trivial for both terminations, as shown in Fig.~\ref{fig16}. In particular, it is complicated by the termination of the all-in-all-out spin structure of the mean-field state, resulting in some $\sigma_z$ component of the spin polarization of Fermi arcs as well as interesting polarization of the in-plane spin components, which are different for the triangular and kagome terminations. This yields very interesting behavior of the quasiparticle interference patterns as discussed in the next section. Interestingly, this turns out to be quite different to the case of the Weyl phases induced by the lack of inversion symmetry\cite{Kourtis2016,Mitchell2016}.

\section{Quasiparticle interference}
\label{surf}

From the Green's function in the slab geometry  we can compute the Green's function in the presence of point-like impurities using the standard $T$-matrix formalism:
\begin{equation}
G_{ij}\left( \omega \right) = G_0^{ij}+\sum_{l,m=1}^N G_0^{il}\left( \omega \right) T_{lm} \left( \omega \right) G_0^{mj} \left( \omega \right),
\label{dyson_exp}
\end{equation}
where the $T$-matrix represents the solution to the scattering problem
\begin{equation}
T_{ij}^{\alpha} \left( \omega \right) = V_{i}^{\alpha} \delta_{i,j} + V_i^{\alpha} \sum^N_{n=1} G_0^{ik} \left( \omega \right) T_{kj}^{\alpha} \left( \omega \right) .
\label{bethe}
\end{equation}
In the case of a point impurity \(V_i^{\alpha}=S^{\alpha}V_0 \delta_{\vec{r}_i,0}\), 
where \(S^{\alpha}=\tau_{i_1} \otimes ... \otimes \tau_{i_n} \otimes \sigma_{\alpha} \) encodes the structure of the impurity in the spin ($\sigma$) and the band ($\tau_i$) basis spaces, respectively. We further assume that the scattering potential and the corresponding $T$-matrix are momentum independent, which yields
\begin{equation}
T^{\alpha} \left( \omega \right) = \left[ 1 - V^{\alpha} \int \frac{ \mathrm d^2 \vec k}{4 \pi^2} G_0\left( \vec k , \omega \right) \right]^{-1} V^{\alpha} .
\end{equation}
As we are interested mostly in the surface states, we consider the position of impurities to be only in the first few layers, close to the termination
\begin{equation}
V_i^{\alpha}=S^{\alpha} V_0 \delta_{r,0} = \sum_{n'}^{N_0=0} S^{\alpha} V_0 \delta_{x,0} \delta_{y,0} \delta_{n_i,n'} \delta_{n',n_j},
\end{equation}
where the limit of the sum indicates the depth through which impurities, still located in the origin of the two-dimensional subsystem, are distributed away from the interface. From the Fourier transform of the Green's function correction in Eq.~(\ref{dyson_exp}) 
\begin{equation}
\begin{split}
&\delta G_{nn'}^{\alpha}\left( \vec k_{||}, \vec q_{||}  ,\omega \right) = \\
&\sum_{n''n'''}^N G_0^{nn''}\left( \vec k_{||} + \vec q_{||} ,\omega \right)  T_{n''n'''}^{\alpha} \left( \omega \right)  G_0^{n'''n'} \left( \vec k_{||}' ,\omega \right)
\end{split}
\end{equation}
we find the spin-resolved correction to the density of states as\cite{queiroz_PRB}
\begin{equation}
\begin{split}
&\delta \rho^{\alpha \beta} \left( \vec q_{||} , \omega \right) = \\
&-\sum_{n_i}^{N_0}\left[ \frac{1}{2 \pi i} \int \frac{\mathrm dk^2}{\left( 2\pi \right)^2} \left[   \mathrm {Tr}\left( S^{\alpha}  \delta G_{n_i,n_i}^{\beta}\left( \vec k_{||},\vec k_{||} + \vec q_{||}, \omega \right) \right) \right. \right. \\
& \left. \left. - \mathrm {Tr}\left( S^{\alpha} \delta G_{n_i,n_i}^{\ast,\beta}\left( \vec k_{||} - \vec q_{||},\vec k_{||}, \omega \right) \right) \right] \right]\\
&=\frac{1}{2\pi i} \left[\Lambda^{\alpha \beta} \left(\vec q_{||},\omega\right)- \Lambda^{\alpha \beta \ast} \left(  - \vec q_{||},\omega \right) \right],
\end{split}
\label{cond_tens}
\end{equation}
where \( \beta \) indicates the spin polarization of the impurity responsible for the scattering, while \(\alpha\) refers to the measurement channel being either charge (\(\alpha=0\)) or spin (\( \alpha=x,y,z\) or \( \alpha=1,2,3\)) ones.

\subsubsection{Non-magnetic point impurity}
We start with a single non-magnetic point impurity on either surface terminations at an energy of \(\omega=0.345 t_o\) which is the energy position of the Weyl crossing for both terminations.

\begin{figure*}[t!]
 \includegraphics[width=1.0\textwidth]{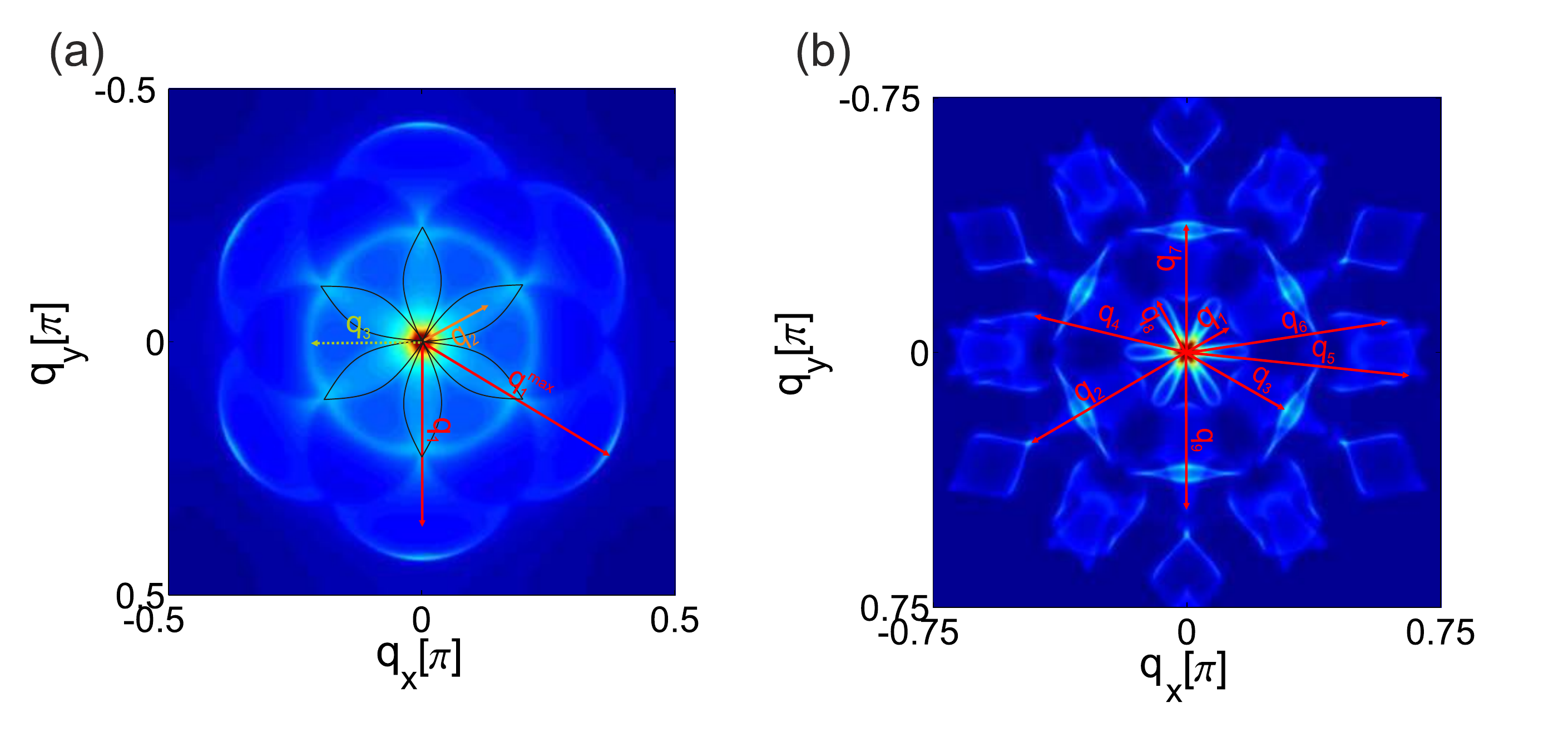}
     \caption{Joint density of the surface states for (a) kagome and (b) triangluar surface termination. The {\bf q}-vectors mark the potential scattering wave vectors, which could be visible in the QPI due to the large joint density of states.}
\label{figJoint}
\end{figure*}

\begin{figure*}[b!]
    \includegraphics[width=1.0\linewidth]{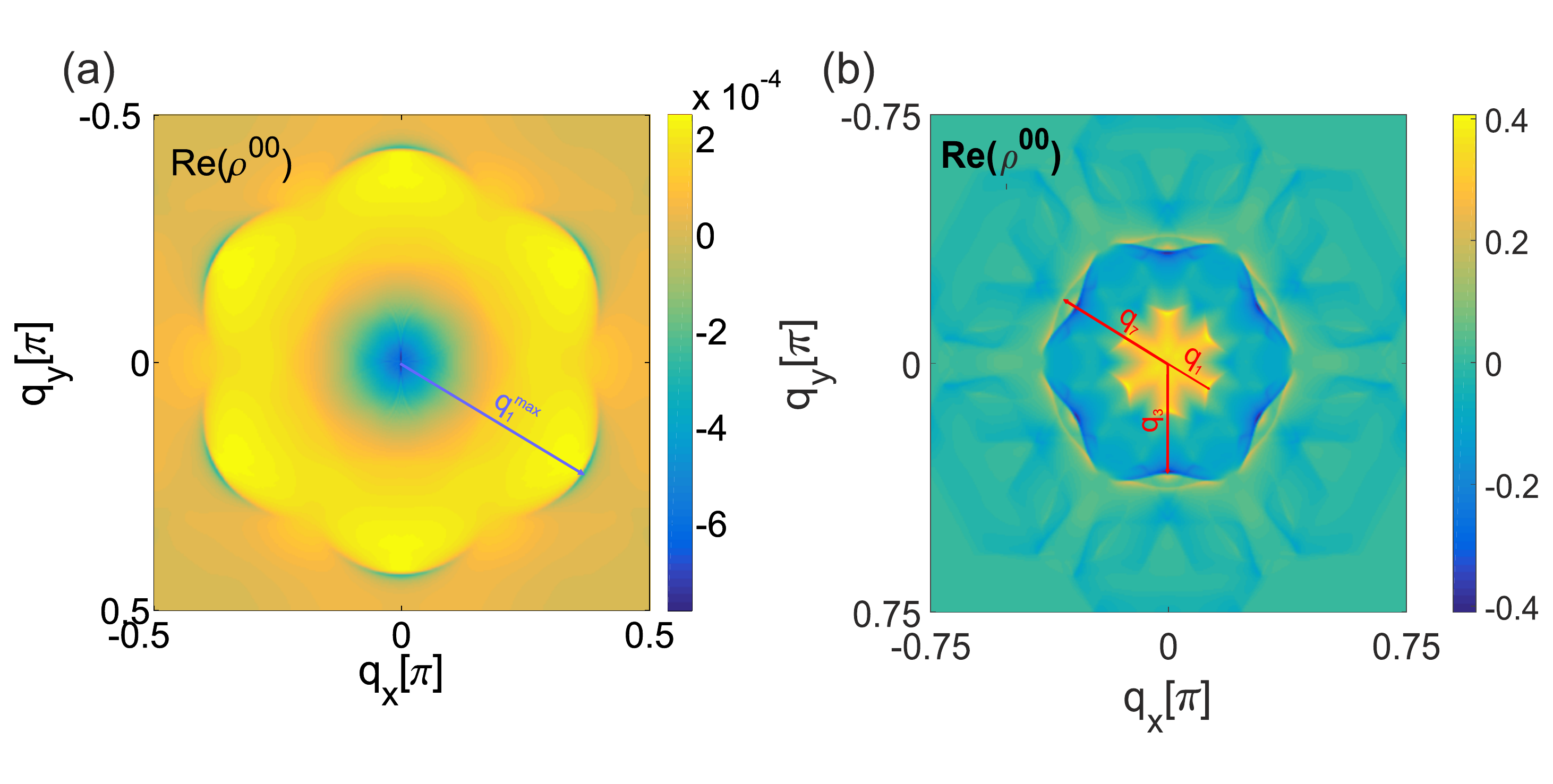}
\caption{The correction to the local density of states (QPI map) in the case of a single nonmagnetic point impurity for kagome (a) and triangular (b) terminations.}
    \label{fig19}
\end{figure*}

Let us first discuss the case of the kagome terminated surface (KTS). 
In this case, shown in Fig.~\ref{fig19}(a), the QPI pattern \(\rho^{00}\) appears to be rather featureless for the KTS. The notable absence of sharp peaks arises from the absence of backscattering. This is a consequence of the electronic structure shown in Fig.~\ref{fig14}, where backscattering partners are absent on one surface. Furthermore, the pattern lacks the ``pinch-point'' at \(\vec q = 0\) that was previously noted to be characteristic for the QPI of Weyl semimetals \cite{Kourtis2016}. In the present case this structure is completely suppressed by the non-trivial spin polarization of the surface states, yet it is perfectly visible in the joint density of states, see Fig.~\ref{figJoint}(a). At the same time, the outer regions of the QPI patterns
clearly show disjoint cross correlation arcs,
which are  clear signatures of the Fermi arcs, see Fig.~\ref{fig19}(a).

\clearpage

Now we turn to the corresponding scattering patterns on the triangular-terminated surface (TTS), see Fig.~\ref{fig19}(b). Here we can see more definite structures in the sense  that we find more distinguishable prominent features in the QPI. The inter-arc cross-correlation on the other hand is far more complicated as the correlations of different pairs appear  in superposition. Although there should exist inter- and intra-arc contributions on the same lenghtscale, the intra-arc correlations are again suppressed by the spin projection of the arcs, as one sees from the joint density of states in Fig.~\ref{figJoint}(b). A signature of the  disconnected nature of the Fermi surface is again the disconnected nature of the scattering pattern itself. It shows six nearly nodal lines which would require a three-fold spin rotation along any closed Fermi-surface with intact spin-momentum locking. This nodal structure can also be further investigated when looking into  QPI patterns in the spin polarized channel.

\begin{figure*}[b!]
\includegraphics[width=0.8\linewidth]{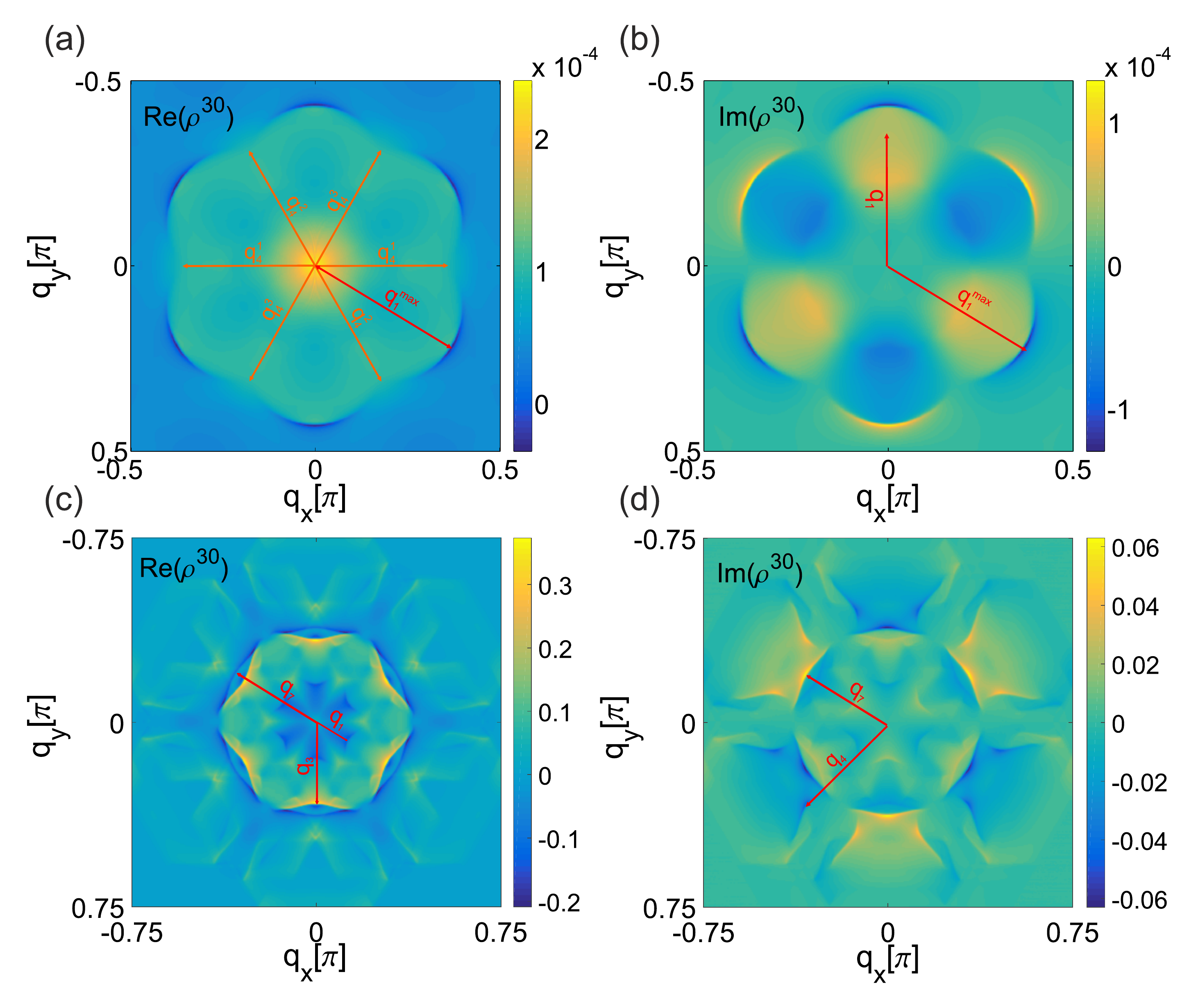}
\caption{The real (left panels) and imaginary (right panels) parts of the Fourier transform of the spin local density of states in the case of single nonmagnetic point impurity for the kagome [(a),(b)] and triangular [(c),(d)] terminations.}
\label{fig20}
\end{figure*}

In particular, we present the real [Fig.~\ref{fig20}(a)] and imaginary [Fig.~\ref{fig20}(b)] parts of the QPI patterns 
in the out-of-plane spin polarized channel for a non-magnetic impurity $(\delta \rho^{30})$. Note that within the Born approximation, due to  symmetry, the QPI patterns in this channel are identical to the QPI patterns in the charge channel for an out-of-plane magnetic impurity $(\delta \rho^{03})$, up to a sign change of $\vec q_{||}$. This can easily be seen by considering the individual terms of Eq.~(\ref{cond_tens}). Exchanging the trace and the integration, the integral can be manipulated as follows
\begin{equation}
\begin{split}
&\Lambda^{\alpha \beta} \left( \vec q_{||} , \omega \right)=\\
 & V_0 \mathrm {Tr} \left[  \int \frac{\mathrm dk^2}{\left( 2\pi \right)^2}
 S^{\alpha} G_0 \left( \vec k_{||} + \vec q_{||} ,\omega \right)  S^{\beta} G_0 \left( \vec k_{||} ,\omega \right)\right]\\
 &= V_0 \mathrm {Tr} \left[  \int \frac{\mathrm dk^{`2}}{\left( 2\pi \right)^2}
 S^{\alpha} G_0 \left( \vec k_{||}' ,\omega \right)  S^{\beta} G_0 \left( \vec k_{||}'-\vec q_{||} ,\omega \right)\right]\\
 &=V_0 \mathrm {Tr} \left[  \int \frac{\mathrm dk^{`2}}{\left( 2\pi \right)^2}
 S^{\beta} G_0 \left( \vec k'_{||} - \vec q_{||} ,\omega \right)  S^{\alpha} G_0 \left( \vec k'_{||} ,\omega \right)\right]\\
 &= \Lambda^{\beta \alpha} \left( -\vec q_{||} , \omega \right).
 \end{split}
\end{equation}
The  real part of \(\rho^{30}\) for the KTS shows essentially the same features as \(\rho^{00}\), which are the cross-correlation arcs of the inter-arc scattering. In comparison to the imaginary part, there is only one additional feature, i.e.,  the finite value of \(\mathbb {Re} \rho^{30}\) at scattering vectors \(\vec q_4\). Since those are wavevectors connecting projected Weyl point positions, that obey inversion symmetry in charge and spin, they have to give no contribution in the imaginary QPI pattern, while they give finite contribution to the real part. Therefore, the  imaginary QPI pattern for the KTS can be considered to be the direct consequence of the non-trivial spin polarization and the arc structure of the surface states. Since the integrand  of \(\mathbb {Im} \rho^{30}\), following Eq.~(\ref{cond_tens}), is proportional to \(\rho_{\vec k }^{x} \rho_{\vec k +\vec q}^y-\rho_{\vec k }^{y} \rho_{\vec k +\vec q}^x \), with \(\rho_{\vec k }^{x} \) and \(\rho_{\vec k }^{x}\) being spin projections of the Greens function, see Eq.~(8), one can see that the absence of parts of the Fermi surface prevent the cancellation of inter-arc scattering. Similar arguments hold for the TTS-surface states, although the completion of the Fermi surface does not correspond to a simple Fermi circle, but a more complicated structure with multiple self-intersections, see Figs.~12(c) and~12(d).

\subsubsection{Classical magnetic point impurity}
\begin{figure*}[h!]
\includegraphics[width=1.0\textwidth]{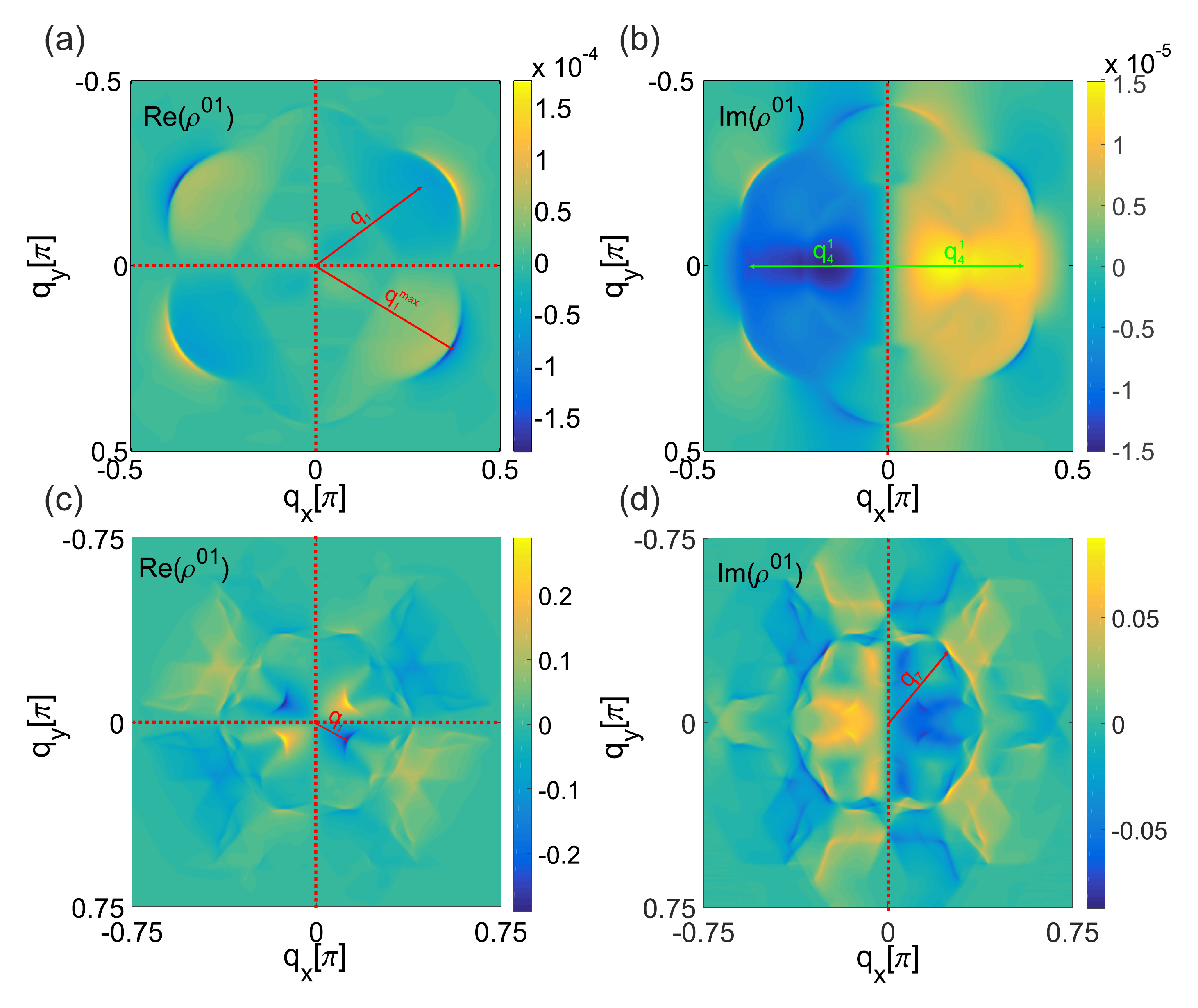}
\caption{The real (left panels) and imaginary (right panels) parts of the Fourier transform of the local density of states in the case of a single magnetic point impurity with  spin polarization in the $x$-direction for the kagome [(a),(b)] and triangular [(c),(d)] terminations.}
\label{fig23}
\end{figure*}

Next, we study the QPI patterns induced by a magnetic impurity polarized along the $x$ direction. Starting with the charge channel of $x$-polarized impurities, we find two nodes along the \(q_x\) and \(q_y\) direction in the real parts of the QPI for both surface terminations,  see Fig.~\ref{fig23}.
Furthermore, information about the in-plane spin polarization symmetries can be gained from the \(\rho^{01}\) pattern. Within the first Born approximation these patterns result from integrating out terms proportional to \((\rho_{\vec k }^{x} \rho_{\vec k +\vec q}^0+\rho_{\vec k }^{0} \rho_{\vec k +\vec q}^x)+i(\rho_{\vec k }^{y} \rho_{\vec k +\vec q}^z-\rho_{\vec k }^{z} \rho_{\vec k +\vec q}^y)\). Therefore the nodes in the real parts are a consequence of the $x$-polarization being antisymmetric under the operation \(k_x  \leftrightarrow -k_x\) and \(k_y  \leftrightarrow -k_y\). The same arguments can be used to explain the \(q_x=0\) node in the imaginary parts as being the consequence of the $y$-polarization symmetry under \(k_y  \leftrightarrow -k_y\), while symmetry under \(k_x  \leftrightarrow -k_x\) cannot be simply deduced from the patterns, due to the lack of a \( q_y=0\) node for both surfaces. Furthermore, it turns out that finding the properties of the $y$-polarization under \(\left( k_x \rightarrow -k_x \right) \) is a little more difficult. The QPI in Figs.~\ref{fig23}(c,d) shows no node in the \(q_x\) direction. The contributions, although small in magnitude compared to other QPI patterns, must result from  antisymmetry with regard to the \(\left( k_x \rightarrow -k_x \right) \) operation of either the $y$- or $z$-polarization, since those show up in combination in the first Born approximation. To decide which one is the case we would have to look at two more QPI patterns, namely \( \mathbb {Im} \rho^{21} \) and  \( \mathbb {Im} \rho^{31} \), which are hard to access in experiment.   \\

\vspace{-0.22cm}

\begin{figure*}[b!]
\includegraphics[width=1.0\linewidth]{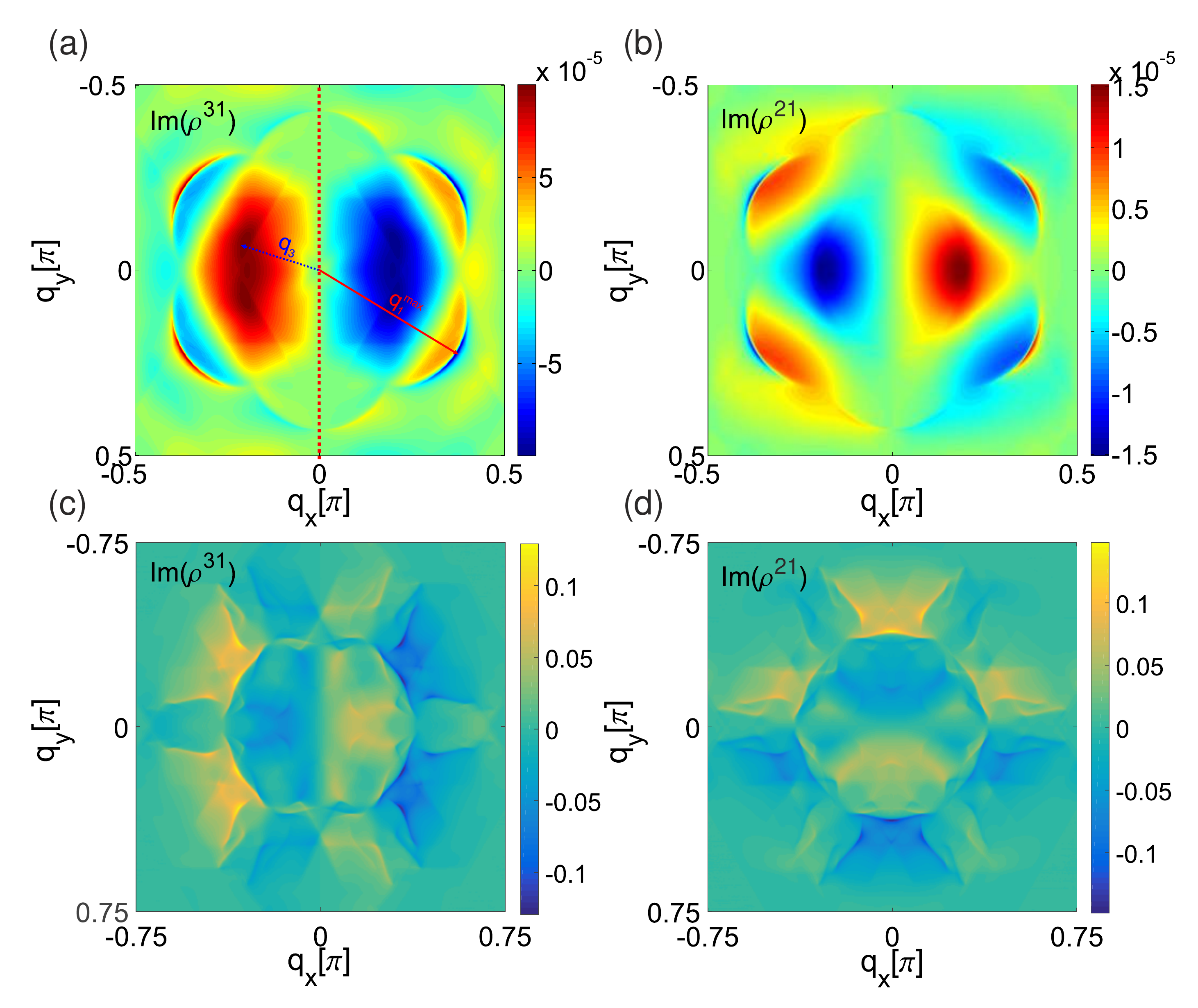}
\caption{The imaginary part of the Fourier transform of the spin-projected density of states for kagome (upper panels) and triangular (lower panels) terminations in the case of a single magnetic point impurity with  spin polarization along  $x$-direction [(a),(c)] and $y$-direction  [(b),(d)].}
\label{fig24}
\end{figure*}

Next, in Figs.~\ref{fig24}(a,c) we show the antisymmetric contributions arising only from the $y$-polarization, since the spectral density is of constant positive sign. Although there is still no node in the \(q_x\) direction, the intensity at \(\vec q_4\) in Fig. \ref{fig24}(a) is close to zero and Fig. \ref{fig24}(c) shows only very weak contributions along that direction. This is in accordance with Fig.\ref{fig24}(b), which includes the $z$-polarized contribution, having peaks at \(\vec q_4\) and Fig. \ref{fig24}(d) having no sign of a node along \(\vec q_x\). We therefore conclude, that the antisymmetric contributions in Figs.\ref{fig23}(c-d) were likely a consequence of antisymmetric behavior of the $z$-polarization under the \(\left( k_x \rightarrow -k_x \right) \) operation.

\clearpage

\subsection{Temperature dependence}

\begin{figure}[]
	\includegraphics[width=0.99\linewidth]{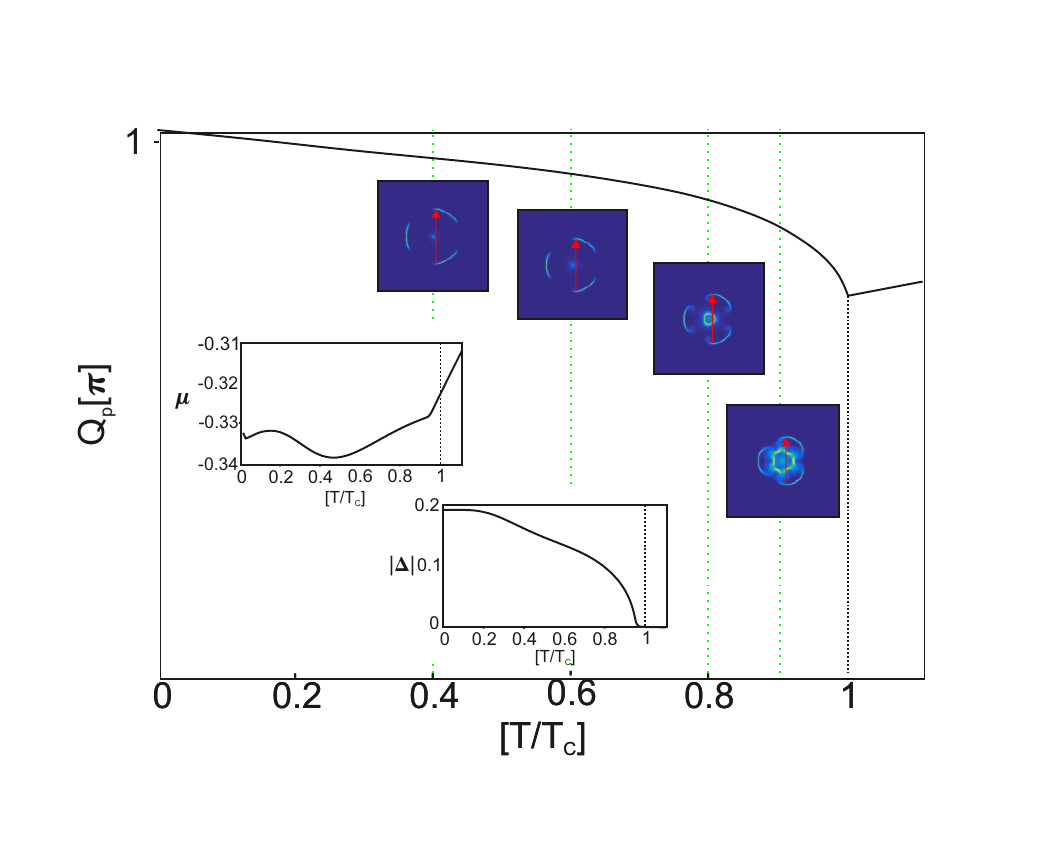}
	\caption{Temperature dependence of the scattering peak together with temperature dependencies of the order parameter $|\Delta|$ (magnetization) and the chemical potential $\mu$.}
	\label{temp_dep}
\end{figure}

Another interesting feature visible in the QPI response of pyrochlore iridates is the temperature dependence of the local magnetization, which  has direct impact on the QPI signal. Since the outer-most features of the QPI originate from scattering between Fermi arcs, the position of these features depend sensitively on the positions of the Weyl points. As both the local magnetization and the chemical potential are temperature dependent, which we capture in a mean-field treatment of the antiferromagnetic phase, the QPI features are bound to be also strongly temperature dependent. Thus both T-dependencies influence the scattering peak position.

The results of our calculations are shown in Fig.~\ref{temp_dep}. One finds that the temperature dependence is not as strong as one might have expected alone from T dependence of the magnetization (lower inset) which should in principle control the Weyl point position. The explanation for this is  the temperature dependence of the chemical potential, which does not follow the energy position of the bulk band crossing of the Weyl point. Basically, at T$>0.9T_C$ one finds not the scattering between Fermi arcs connecting Weyl point projections, but between some surface (non-Weyl) states. These are located well above the energy position of the Weyl crossing which occurs at higher values of \(\vec q\). 
Nonetheless there is an observable temperature dependence of a quite isolated scattering peak for some temperature region below $T_c$, connected to the dispersion of the Fermi arc. \\
\section{Conclusion}
\label{conc}
To conclude, in this paper we analyzed the QPI signatures of the Fermi arcs in time-reversal breaking Weyl semimetals.
As a concrete example, we considered the Weyl phase of an  interacting multi-band Hubbard model on the pyrochlore
lattice with antiferromangetic spin configuration, as realized in R$_2$Ir$_2$O$_7$.  We demonstrated that the use of various types of impurities allows not only to prove the  very existence of the Fermi arcs but also to reconstruct from the QPI patterns the unique features of the surface states and their internal spin polarization. This can help to identify the wave function of the surface fermions and their spin-momentum locking. Furthermore, we find that for Weyl phases induced by many-body effects, the scattering features associated with the Fermi arcs show a very strong temperature dependence that stems from the many-body origin of the Weyl phase. 

\acknowledgments

We would like to thank Alireza Akbari, Flavio Nogueira, and Andreas Rost for very useful discussions. This work was supported by the DFG Schwerpunkt Program "Topologische Isolatoren" (ER 463/9-1) and SFB 1143 "Korrelierter Magnetismus: Von Frustration zu Topologie".


\newpage
\newpage

\appendix

\begin{widetext}

\section{Dimensional reduction matrices}
\label{H_s}

The Hamiltonian in the slab-geometry, given by Eq.(\ref{slab}) in the main text, has the following elements. In particular, one finds for the nearest neighbour terms:
\begin{equation}
\begin{split}
H^{0,NN}_k&= 2
\begin{pmatrix}
0 & \Gamma_{0,1,-1}^{+}  \cos \left( \frac{3}{\sqrt{6}}k_y+ \frac{1}{\sqrt{2}}k_x \right) & \Gamma_{-1,0,1}^{+}\cos \left( \sqrt{2}k_x \right) \\
\Gamma_{0,1,-1}^{-} \cos \left( \frac{3}{\sqrt{6}}k_y+ \frac{1}{\sqrt{2}}k_x \right) & 0 & \Gamma_{1,1,0}^{+} \cos \left( \frac{1}{\sqrt{2}}k_x- \frac{3}{\sqrt{6}}k_y \right) \\
\Gamma_{-1,0,1}^{-}\cos \left( \sqrt{2}k_x \right)  & \Gamma_{1,1,0}^{-} \cos \left( \frac{1}{\sqrt{2}}k_x- \frac{3}{\sqrt{6}}k_y \right) & 0
\end{pmatrix}\\
H^{0,NN}_t&=0\\
H^{\pm1,NN}_{k-t}&=
\begin{pmatrix}
 \Gamma_{-1,1,0}^{+}\left(\cos \left( \frac{1}{\sqrt{2}}k_x+\frac{1}{\sqrt{6}}k_y \right) \pm i\sin \left( \frac{1}{\sqrt{2}}k_x+\frac{1}{\sqrt{6}}k_y \right) \right)  \\
\Gamma_{-1,0,-1}^{+}\left( \cos \left( \frac{2}{\sqrt{6}}k_y \right) \mp i\sin \left( \frac{2}{\sqrt{6}}k_y \right) \right) \sigma_0\\
\Gamma_{0,1,1}^{+}\left( \cos \left( \frac{1}{\sqrt{2}}k_x-\frac{1}{\sqrt{6}}k_y \right) \pm i\sin \left( \frac{1}{\sqrt{2}}k_x-\frac{1}{\sqrt{6}}k_y \right) \right) \sigma_0
\end{pmatrix}\\
H^{\pm1,NN}_{t-k}&=
\begin{pmatrix}
 \Gamma_{-1,1,0}^{-}\left(\cos \left( \frac{1}{\sqrt{2}}k_x+\frac{1}{\sqrt{6}}k_y \right) \pm i\sin \left( \frac{1}{\sqrt{2}}k_x+\frac{1}{\sqrt{6}}k_y \right) \right)  \\
\Gamma_{-1,0,-1}^{-}\left( \cos \left( \frac{2}{\sqrt{6}}k_y \right) \mp i\sin \left( \frac{2}{\sqrt{6}}k_y \right) \right) \sigma_0\\
\Gamma_{0,1,1}^{-}\left( \cos \left( \frac{1}{\sqrt{2}}k_x-\frac{1}{\sqrt{6}}k_y \right) \pm i\sin \left( \frac{1}{\sqrt{2}}k_x-\frac{1}{\sqrt{6}}k_y \right) \right) \sigma_0
\end{pmatrix}^{T}\\
H^{\pm i,NN}_{\left(k-t\right),\left(t-k\right)}&=0, if \ |i|>1\\
\end{split}
\end{equation}
where the abbreviation \( \Gamma_{b,c,d}^{\pm}= \left(  t_1\sigma_0 \pm it_2 \left( b \sigma_x + c \sigma_y + d \sigma_z \right) \right) \) was used.
The next-nearest-neighbour hopping terms could be written as
\begin{equation}
\begin{split}
\prescript{}{11}H^{0,NNN}_k&=0\\
\prescript{}{12}H^{0,NNN}_k&=2 \left(  t_1' \sigma_0+ i t_2' \Pi_{--+} + i t_3' \Pi_{+-+} \right) \cos \left( \frac{3}{\sqrt{2}} k_x - \frac{3}{\sqrt{6}} k_y \right)\\
\prescript{}{13}H^{0,NNN}_k&=2 \left(  t_1' \sigma_0+ i t_2' \Pi_{++-} + i t_3' \Pi_{+--} \right) \cos \left( \sqrt{6} k_y \right)\\
\prescript{}{21}H^{0,NNN}_k&= \prescript{}{12}H^{0,NNN,\dagger}_k\\
\prescript{}{22}H^{0,NNN}_k&=0\\
\prescript{}{23}H^{0,NNN}_k&= 2\left(  t_1' \sigma_0+ i t_2' \Pi_{--+} + i t_3' \Pi_{---} \right) \cos \left(\frac{3}{\sqrt{2}} k_x + \frac{3}{\sqrt{6}} k_y  \right)\\
\prescript{}{31}H^{0,NNN}_k&= \prescript{}{13}H^{0,NNN,\dagger}_k\\
\prescript{}{32}H^{0,NNN}_k&= \prescript{}{23}H^{0,NNN,\dagger}_k\\
\prescript{}{33}H^{0,NNN}_k&=0\\
\prescript{}{ij}H^{0,NNN}_t&=0\\
\end{split}
\end{equation}
\begin{equation}
\begin{split}
\prescript{}{11}H^{\pm1,NNN}_{k-t}&= \left(  t_1' \sigma_0+ i t_2' \Pi_{-+-} + i t_3' \Pi_{-++} \right) \left( \cos \left(-\frac{1}{\sqrt{2}} k_x - \frac{5}{\sqrt{6}} k_y  \right) \pm i \sin \left(-\frac{1}{\sqrt{2}} k_x - \frac{5}{\sqrt{6}} k_y  \right) \right)+\\
&\left(  t_1' \sigma_0+ i t_2' \Pi_{-++} + i t_3' \Pi_{-+-} \right) \left( \cos \left(-\frac{3}{\sqrt{2}} k_x + \frac{1}{\sqrt{6}} k_y  \right) \pm i \sin \left(-\frac{3}{\sqrt{2}} k_x + \frac{1}{\sqrt{6}} k_y  \right) \right)\\
\prescript{}{21}H^{\pm1,NNN}_{k-t}&= \left(  t_1' \sigma_0+ i t_2' \Pi_{+-+} + i t_3' \Pi_{+++} \right) \left( \cos \left(\frac{2}{\sqrt{2}} k_x + \frac{4}{\sqrt{6}} k_y  \right) \pm i \sin \left(\frac{2}{\sqrt{2}} k_x + \frac{4}{\sqrt{6}} k_y  \right) \right)+\\
&\left(  t_1' \sigma_0+ i t_2' \Pi_{+++} + i t_3' \Pi_{+-+} \right) \left( \cos \left(-\frac{2}{\sqrt{2}} k_x + \frac{4}{\sqrt{6}} k_y  \right) \pm i \sin \left(-\frac{2}{\sqrt{2}} k_x + \frac{4}{\sqrt{6}} k_y  \right) \right)\\
\prescript{}{31}H^{\pm1,NNN}_{k-t}&= \left(  t_1' \sigma_0+ i t_2' \Pi_{+--} + i t_3' \Pi_{---} \right) \left( \cos \left(\frac{3}{\sqrt{2}} k_x + \frac{1}{\sqrt{6}} k_y  \right) \pm i \sin \left(\frac{3}{\sqrt{2}} k_x + \frac{1}{\sqrt{6}} k_y  \right) \right)+\\
&\left(  t_1' \sigma_0+ i t_2' \Pi_{---} + i t_3' \Pi_{+--} \right) \left( \cos \left(\frac{1}{\sqrt{2}} k_x - \frac{5}{\sqrt{6}} k_y  \right) \pm i \sin \left(\frac{1}{\sqrt{2}} k_x + \frac{5}{\sqrt{6}} k_y  \right) \right)\\
\prescript{}{11}H^{\pm1,NNN}_{t-k}&= \left(  t_1' \sigma_0+ i t_2' \Pi_{-+-} + i t_3' \Pi_{-++} \right)^{\dagger} \left( \cos \left(-\frac{1}{\sqrt{2}} k_x - \frac{5}{\sqrt{6}} k_y  \right) \pm i \sin \left(-\frac{1}{\sqrt{2}} k_x - \frac{5}{\sqrt{6}} k_y  \right) \right)+\\
&\left(  t_1' \sigma_0+ i t_2' \Pi_{-++} + i t_3' \Pi_{-+-} \right)^{\dagger} \left( \cos \left(-\frac{3}{\sqrt{2}} k_x + \frac{1}{\sqrt{6}} k_y  \right) \pm i \sin \left(-\frac{3}{\sqrt{2}} k_x + \frac{1}{\sqrt{6}} k_y  \right) \right)\\
\prescript{}{12}H^{\pm1,NNN}_{t-k}&= \left(  t_1' \sigma_0+ i t_2' \Pi_{+-+} + i t_3' \Pi_{+++} \right)^{\dagger} \left( \cos \left(\frac{2}{\sqrt{2}} k_x + \frac{4}{\sqrt{6}} k_y  \right) \pm i \sin \left(\frac{2}{\sqrt{2}} k_x + \frac{4}{\sqrt{6}} k_y  \right) \right)+\\
&\left(  t_1' \sigma_0+ i t_2' \Pi_{+++} + i t_3' \Pi_{+-+} \right)^{\dagger} \left( \cos \left(-\frac{2}{\sqrt{2}} k_x + \frac{4}{\sqrt{6}} k_y  \right) \pm i \sin \left(-\frac{2}{\sqrt{2}} k_x + \frac{4}{\sqrt{6}} k_y  \right) \right)\\
\prescript{}{13}H^{\pm1,NNN}_{t-k}&= \left(  t_1' \sigma_0+ i t_2' \Pi_{+--} + i t_3' \Pi_{---} \right)^{\dagger} \left( \cos \left(\frac{3}{\sqrt{2}} k_x + \frac{1}{\sqrt{6}} k_y  \right) \pm i \sin \left(\frac{3}{\sqrt{2}} k_x + \frac{1}{\sqrt{6}} k_y  \right) \right)+\\
&\left(  t_1' \sigma_0+ i t_2' \Pi_{---} + i t_3' \Pi_{+--} \right)^{\dagger} \left( \cos \left(\frac{1}{\sqrt{2}} k_x - \frac{5}{\sqrt{6}} k_y  \right) \pm i \sin \left(\frac{1}{\sqrt{2}} k_x + \frac{5}{\sqrt{6}} k_y  \right) \right)\\
\prescript{}{11}H^{\pm2,NNN}_{k-k}&=0\\
\prescript{}{12}H^{\pm2,NNN}_{k-k}&= \left(  t_1' \sigma_0+ i t_2' \Pi_{+-+} + i t_3' \Pi_{--+} \right) \left(\cos \left(-\frac{1}{\sqrt{2}} k_x + \frac{1}{\sqrt{6}} k_y  \right) \mp i\sin \left(-\frac{1}{\sqrt{2}} k_x + \frac{1}{\sqrt{6}} k_y  \right) \right) \\
\prescript{}{13}H^{\pm2,NNN}_{k-k}&= \left(  t_1' \sigma_0+ i t_2' \Pi_{+--} + i t_3' \Pi_{++-} \right) \\
\prescript{}{21}H^{\pm2,NNN}_{k-k}&= \left(  t_1' \sigma_0+ i t_2' \Pi_{+-+} + i t_3' \Pi_{--+} \right)^{\dagger} \left(\cos \left(-\frac{1}{\sqrt{2}} k_x + \frac{1}{\sqrt{6}} k_y  \right) \mp i\sin \left(-\frac{1}{\sqrt{2}} k_x + \frac{1}{\sqrt{6}} k_y  \right) \right) \\
\prescript{}{22}H^{\pm2,NNN}_{k-k}&=0\\
\prescript{}{23}H^{\pm2,NNN}_{k-k}&= \left(  t_1' \sigma_0+ i t_2' \Pi_{---} + i t_3' \Pi_{--+} \right) \left(\cos \left(-\frac{1}{\sqrt{2}} k_x + \frac{1}{\sqrt{6}} k_y  \right) \mp i\sin \left(-\frac{1}{\sqrt{2}} k_x + \frac{1}{\sqrt{6}} k_y  \right) \right) \\
\prescript{}{31}H^{\pm2,NNN}_{k-k}&= \left(  t_1' \sigma_0+ i t_2' \Pi_{+--} + i t_3' \Pi_{++-} \right)^{\dagger} \\
\prescript{}{32}H^{\pm2,NNN}_{k-k}&= \left(  t_1' \sigma_0+ i t_2' \Pi_{---} + i t_3' \Pi_{--+} \right)^{\dagger} \left(\cos \left(-\frac{1}{\sqrt{2}} k_x + \frac{1}{\sqrt{6}} k_y  \right) \mp i\sin \left(-\frac{1}{\sqrt{2}} k_x + \frac{1}{\sqrt{6}} k_y  \right) \right) \\
\prescript{}{33}H^{\pm2,NNN}_{k-k}&=0\\
\prescript{}{ij}H^{\pm2,NNN}_{t-t}&=0, \ \forall i,j
\end{split}
\end{equation}
where \( \Pi_{\pm \pm \pm}= \left( \pm \sigma_x \pm \sigma_y \pm \sigma_z \right) \).
Finally the mean field interaction terms in this representation are given by:
\begin{equation}
\begin{split}
H^{int}_k  &= \frac{U \Delta}{\sqrt{3}}
\begin{pmatrix}
\Pi_{---} & 0 & 0\\
0 & \pi_{-++} & 0\\
0 & 0 & \Pi_{+-+}\\
\end{pmatrix}\\
H^{int}_t  &=\frac{U \Delta}{\sqrt{3}} \Pi_{++-}\\
\end{split}
\end{equation}
The complete submatrices will therefore have the following form:
\begin{equation}
\begin{split}
H^0_k  &= H^{0,NN}+H^{int}_k +
\begin{pmatrix}
\prescript{}{11}H^{0,NNN}_k & \prescript{}{12}H^{0,NNN}_k & \prescript{}{13}H^{0,NNN}_k\\
\prescript{}{21}H^{0,NNN}_k & \prescript{}{22}H^{0,NNN}_k & \prescript{}{23}H^{0,NNN}_k\\
\prescript{}{31}H^{0,NNN}_k & \prescript{}{32}H^{0,NNN}_k & \prescript{}{33}H^{0,NNN}_k\\
\end{pmatrix}\\
H^0_t  &= H^{int}_t\\
H^{\pm1}_{k-t}&=H^{\pm1,NN}_{k-t}+
\begin{pmatrix}
\prescript{}{11}H^{\pm1,NNN}_{k-t}\\
\prescript{}{21}H^{\pm1,NNN}_{k-t}\\
\prescript{}{31}H^{\pm1,NNN}_{k-t}\\
\end{pmatrix}\\
H^{\pm1}_{t-k}&=H^{\pm1,NN}_{t-k}+
\begin{pmatrix}
\prescript{}{11}H^{\pm1,NNN}_{t-k} & \prescript{}{12}H^{\pm1,NNN}_{t-k} & \prescript{}{13}H^{\pm1,NNN}_{t-k}\\
\end{pmatrix}\\
H^{\pm2}_{k-k}&=
\begin{pmatrix}
\prescript{}{11}H^{\pm2,NNN}_{k-k} & \prescript{}{12}H^{\pm2,NNN}_{k-k} & \prescript{}{13}H^{\pm2,NNN}_{k-k}\\
\prescript{}{21}H^{\pm2,NNN}_{k-k} & \prescript{}{22}H^{\pm2,NNN}_{k-k} & \prescript{}{23}H^{\pm2,NNN}_{k-k}\\
\prescript{}{31}H^{\pm2,NNN}_{k-k} & \prescript{}{32}H^{\pm2,NNN}_{k-k} & \prescript{}{33}H^{\pm2,NNN}_{k-k}\\
\end{pmatrix}\\
H^{\pm2}_{t-t}&=0\\
\end{split}
\end{equation}

\end{widetext}

\end{document}